\documentclass[aps,prd,nofootinbib,superscriptaddress,onecolumn,notitlepage,preprintnumbers]{revtex4-1}

\usepackage{amsmath,graphicx,verbatim,epsfig}
\usepackage{epstopdf}
\usepackage[utf8x]{inputenc}
\usepackage{amsmath,graphicx}
\usepackage[normalem]{ulem}
\usepackage{datetime}
\usepackage[colorlinks=true,linkcolor=blue,citecolor=blue]{hyperref}
\usepackage{slashed}

\usepackage[table]{xcolor}
\usepackage[textsize=scriptsize,textwidth=1.5cm]{todonotes}
\usepackage{setspace}

\newcommand{\eps}{\varepsilon}

\newcommand{\nn}{\nonumber}

\newcommand{\bn}{{\bar n}}

\newcommand{\pslash}{{\not \!p}}
\newcommand{\kslash}{{\not \!k}}
\newcommand{\nslash}{{\not \!n}}

\newcommand{\bnslash}{{\not \!\bn}}

\newcommand{\nb}{\bar n}

\newcommand{\veuv}{\varepsilon_{\rm {UV}}}

\newcommand{\be}{\begin{equation}}
\newcommand{\ee}{\end{equation}}
\newcommand{\bea}{\begin{eqnarray}}
\newcommand{\eea}{\end{eqnarray}}
\newcommand{\balign}{\begin{align}}
\newcommand{\ealign}{\end{align}}
\newcommand{\as}{\alpha_s}

\newcommand{\cd}{\cdot}

\newcommand{\sandwich}[3]{\left< #1 \right | #2 \left | #3 \right >}

\newcommand{\bg}{\begin{gather}}
\newcommand{\foma}{\end{gather}}

\newcommand{\noopsort}[1]{}

\newcommand{\vecb}[1]{\mbox{\boldmath $#1$}}
\newcommand{\vecbe}[1]{\mbox{\boldmath ${\scriptstyle #1}$}}
\newcommand{\vecbp}[1]{\mbox{\boldmath $#1_\perp$}}

\def\e{\epsilon}

\def\ve{\varepsilon}

\def\pd{\partial}

\def\L{\Lambda}

\def\z{\zeta}

\def\S{\Sigma}

\def\<{\langle}
\def\>{\rangle}

\def\a{\alpha}

\def\g{\gamma}  \def\G{\Gamma}
\def\d{\delta}  \def\D{\Delta}
\def\l{\lambda}   \def\L{\Lambda}
\def\s{\sigma}
\def\r{\rho}  
\def\x{\xi}

\def\m{\mu}
\def\n{\nu}
\def\t{\tau}

\def\z{\zeta}

\def\({\left(}
\def\[{\left[}
\def\){\right)}
\def\]{\right]}

\def\ln{\hbox{ln}}
\def\log{\hbox{log}}

\def\Qslash{Q\!\!\!\!\slash}

\def\nslash{n\!\!\!\slash}
\def\bnslash{\bar n\!\!\!\slash}
\def\pslash{p\!\!\!\slash}
\def\bpslash{\bar p\!\!\!\slash}

\def\kslash{k\!\!\!\slash}

\def \le { \left }
\def \ri { \right}
\def\bp{\bar p}
\def\bP{\bar P}

\def\lqcd{\L_{\rm QCD}}

\newcommand{\ben}{\begin{eqnarray}}
\newcommand{\een}{\end{eqnarray}}

\newcommand{\bef}{\begin{figure}[htb]\centering}
\newcommand{\eef}{\end{figure}}

\def\avpht{\langle {\hat P}_{hT}^2\rangle}

\begin{document}

\title{A Unified Treatment of the QCD Evolution of All (Un-)Polarized TMD Functions:\\ 
Collins Function as a Study Case}

\author{Miguel G. Echevarria}\email{m.g.echevarria@nikhef.nl}
\affiliation{Nikhef Theory Group,
Science Park 105, 1098 XG Amsterdam, The Netherlands}
\author{Ahmad Idilbi}\email{aui13@psu.edu}
\affiliation{Department of Physics, Pennsylvania State University, 
University Park, PA 16802, USA}
\author{Ignazio Scimemi}\email{ignazios@fis.ucm.es}
\affiliation{Departamento de F\'isica Te\'orica II,
Universidad Complutense de Madrid,
28040 Madrid, Spain}

\date{\today}

\begin{abstract}

By considering semi-inclusive deep-inelastic scattering and the 
(complementary) $q_T$-spectrum for Drell-Yan lepton pair production we derive the QCD evolution for all the leading-twist transverse momentum dependent distribution and fragmentation functions. 
We argue that all of those functions evolve with $Q^2$ following a \emph{single} evolution kernel. 
This kernel is independent of the underlying kinematics and it is also spin-independent. 
Those features hold, in impact parameter space, to all values of $b_T$. 
The evolution kernel presented has all of its large logarithms resummed up next-to-next-to leading logarithmic accuracy, which is the highest possible accuracy given existing perturbative calculations. 
As a study case we apply this kernel to investigate the evolution of Collins function, one of the ingredients that have attracted recently much attention within the phenomenological studies of spin asymmetries.
Our analysis can be readily implemented to revisit previously obtained fits that involve data at different scales for other spin-dependent functions. 
Such improved fits are important to get better predictions --with the correct evolution kernel-- for certain upcoming experiments aiming to measure Sivers function, Collins function, transversity and other spin-dependent functions as well.

\end{abstract}

\preprint{NIKHEF 2014-006}
\maketitle

\section{Introduction}
\label{sec:intro}

Hadronic matrix elements with transverse-momentum dependence (TMD) are indispensable quantities in current high-energy phenomenology. 
Ranging from the LHC physics to the study of the spin and the three-dimensional structure of nucleons, the role of those matrix elements is a footprint of the QCD dynamics.  
In this work we focus on matrix elements that acquire transverse momentum dependence of partons inside the colliding and/or emerging nucleons in deep-inelastic scattering (DIS) type of experiments and also for Drell-Yan (DY) heavy lepton pair production. 
For initial and final state hadronic matrix elements we refer to them below as TMD parton distribution functions (TMDPDFs) and TMD fragmentation functions (TMDFFs), respectively.
Collectively we call them ``TMDs''. 
In semi-inclusive DIS (SIDIS) experiments or the DY $q_T$-dependent spectrum, different TMDs contribute, at leading-twist, to the factorization of the QCD hadronic tensor depending on the polarization of the involved hadrons/partons.
In order to study the hadronic spin structure one needs to consider polarized and/or unpolarized hadrons or partons; thus one needs to define quantities that are sensitive to different polarizations of partons inside polarized or unpolarized hadrons. 
When SIDIS and DY processes are considered, and based on leading-twist factorization theorems and different spin projections of the relevant hadronic tensors for those two processes, one obtains sixteen different TMDs~\cite{Boer:1997nt}. 
Eight of them are related to initial state hadronic matrix elements, and the other eight to final state ones.

It has been well-known for long time that the TMDs acquire, on top of the usual renormalization/factorization scale dependence, an additional  $Q^2$-dependence, where $Q^2$ is the hard probe in a typical high-energy reaction. The last observation does not apply to the integrated (or ``collinear'') parton distribution or fragmentation functions. 
This difference makes the study of the TMDs more interesting since, among other things, one needs to consider the QCD evolution of such quantities with respect to a second scale, namely $Q$, on top of the standard renormalization scale $\mu$.
This $Q^2$-dependence appears at the intermediate scale $q_T$ in an ``anomalous'' manner. 
The ``anomaly'' is that there are, at the intermediate scale, two types~\footnote{This is unlike the case for inclusive observables where, at the intermediate scale --assuming, e.g., partonic threshold-- there is only one type of large logarithms. For DIS one has $\ln(Q^2(1-x))$ and for DY there is $\ln(Q^2(1-z)^2)$.}  of large logarithms that need to be resummed: 
$\ln(Q^2/\mu^2)$ and $\ln(q_T^2/\mu^2)$. 
Needless to say that this extra $Q^2$-dependence results from the fact that the relevant observables are sensitive to the partonic transverse momentum inside the colliding or emerging hadrons. 
As such, this $Q^2$-dependence serves to unravel both the momentum distribution of partons inside hadrons and the fragmentation process of partons to hadrons, where both aspects are complementary to each other and are fundamental to understand certain aspects of QCD dynamics.

In this work we focus on the evolution of all those spin-dependent and independent TMDs with respect to $Q^2$.
In order to do so one needs first to properly define them. 
As we argue below, the role of the soft function(s) (and its splitting thereof) is crucial to obtain well-defined TMDs, and this will ultimately determine the QCD evolution properties of TMD functions. 
In this sense, we generalize the results of Refs.~\cite{Collins:2011zzd,Echevarria:2012js} from the case of unpolarized TMDs to the spin-dependent ones, and from TMDPDFs to TMDFFs.
The issue of evolution of different specific TMDs has received much attention lately and for different TMDs (see e.g.~\cite{Aybat:2011zv,Aybat:2011ta,Aybat:2011ge,Sun:2013hua,Sun:2013dya,Boer:2013zca,Anselmino:2013lza,Anselmino:2013rya,Anselmino:2013vqa,Bacchetta:2013pqa,Aidala:2014hva,Echevarria:2014xaa,Echevarria:2012pw}) and it is of much relevance to HERMES, COMPASS, JLab, Belle, RHIC and LHC.

In Ref.~\cite{GarciaEchevarria:2011rb} we considered the unpolarized DY $q_T$-spectrum for small $q_T$, and showed that the hadronic tensor factorizes into hard, soft and two (pure) collinear matrix elements. 
The factorization theorem then allowed us to combine the collinear contributions with the relevant part of the soft function~\cite{GarciaEchevarria:2011rb,Echevarria:2012js} in order to cancel rapidity divergences. 
The resulting quantity was defined as the unpolarized TMDPDF. 
Through the factorization theorem we also obtained the evolution kernel of the unpolarized quark TMDPDF.
In Ref.~\cite{Echevarria:2012pw} we obtained a resummed evolution kernel where all the large logarithms were resummed up to next-to-next-to leading logarithmic (NNLL) accuracy (N$^3$LL expressions were also provided).
Although obtained from the unpolarized DY hadronic tensor, we argued in Ref.~\cite{Echevarria:2012pw} that the evolution kernel is spin-independent and universal among all initial state-state TMDPDFs, and we applied it to Sivers function.

In going from  initial-state hadronic matrix elements, TMDPDFs, to final-state ones, TMDFFs, we consider the latter ones via SIDIS process.
After deriving its factorization theorem by using soft-collinear effective theory (SCET)~\cite{Bauer:2000yr,Bauer:2001yt,Beneke:2002ph,Bauer:2001ct},
we properly define polarized and unpolarized TMDFFs and obtain their evolution kernel, while resumming large logarithms to NNLL accuracy. 
Similar to the evolution kernel of the TMDPDFs, the resummed  kernel for TMDFFs is spin-independent, and thus it applies to all eight functions which are dependent on final state hadronic matrix elements.
Moreover, by considering some novel features of the soft function (whether the one relevant for SIDIS or DY kinematics)~\footnote{The soft functions appearing in the factorization theorems of DY and SIDIS differ, at operator level, by different structure of the soft Wilson lines.} which enters into the definition of all the sixteen TMDs, we argue that it is universal. This is a major step in establishing that the evolution kernel of all TMDFFs is exactly the same as the one of the TMDPDFs. In other words, all of the sixteen TMDs evolve according to a \emph{single} evolution kernel.~\footnote{In Ref.~\cite{Idilbi:2004vb} (see also Ref.~\cite{Sun:2013hua}) the universality and spin-independence of the Collins-Soper evolution kernel was considered for TMDPDFs, but not for TMDFFs. Moreover, and more importantly, the TMDs considered there have different definitions than the ones introduced in this work regarding the role of the relevant soft function contributions.} 
This fact has a rather important phenomenological implications as we discuss below.

This paper is organized as follows. 
In Sec.~\ref{sec:factth} we derive a factorization theorem for SIDIS using effective field theory methodology and we properly define, by taking into account the soft function contributions and after spin decompositions, all the sixteen relevant TMDs at leading-twist.  
In Sec.~\ref{sec:evolutiontmds} we discuss the evolution of the newly defined TMDs and discuss the universality and the spin-(non)dependence of the evolution kernel. 
In Sec.~\ref{sec:collins} we apply the evolution kernel (after resummation) to Collins function as a study case, representing any of the TMDFFs.
In Appendix~\ref{app:tmdff_nlo} we explicitly calculate the unpolarized TMDFF to ${\cal O}(\as)$ and we show, as expected, that when the pure collinear contribution is combined with the proper soft contribution, all rapidity divergences cancel out. 
In Appendix~\ref{sec:tmdffope} we perform the matching of the unpolarized TMDFF onto the collinear fragmentation function and obtain the Wilson coefficient (which is free from any infra-red/rapidity divergence regulator while all calculations are performed on the-light-cone).  
As a trivial check, in Appendix~\ref{app:hard} we utilize the results of Appendix~\ref{app:tmdff_nlo}  to obtain the hard part relevant for SIDIS kinematics (which has to be the same one as for the inclusive DIS).

\section{Factorization Theorem and Definitions}
\label{sec:factth}

In Ref.~\cite{GarciaEchevarria:2011rb} we derived a factorization theorem for small-$q_T$ DY lepton pair production, highlighting the role of the soft gluon radiation through a well-defined soft function. 
In this section we follow the same steps and derive a factorization theorem for SIDIS case:
\begin{align}
l(k)+N(P) \to l'(k')+h(P_h)+X(P_X)
\,,
\end{align}
where $l$($l'$) is the incoming (outgoing) lepton, $N$ is the nucleon and $h$ is the detected hadron, for which we measure its transverse momentum.
This process is commonly described in terms of the following Lorentz invariants,
\begin{align}
x_B = \frac{Q^2}{2P\cd q}\,,
\quad\quad
y = \frac{P\cd q}{P\cd l}\,,
\quad\quad
z_h = \frac{P\cd P_h}{P\cd q}
\,.
\end{align}
The photon carries momentum $q=k-k'$ with $q^2=-Q^2$.
In the Breit frame, the incoming nucleon $N$ is traveling along the $+z$-direction, with $n$-collinear momentum $P$, and the photon is $\bn$-collinear, traveling along the $-z$-direction~\footnote{A generic vector $v^\m$ is decomposed as 
$v^\m=\bn\cd v\frac{n^\m}{2}+n\cd v\frac{\bn^\m}{2}+v_\perp^\m=
(\bn\cd v, n\cd v, v_\perp^\m)=
(v^+,v^-,v_\perp^\m)$, with $n=(1,0,0,1)$, $\bn=(1,0,0,-1)$, $n^2=\bn^2=0$ and $n\cd\bn=2$. We also use $v_T=|\vecb v_\perp|$, so that $v_\perp^2=-v_T^2$.}.
The outgoing hadron $h$ has a momentum $P_h$ mainly along the $-z$-direction, acquiring a transverse momentum $\vecb P_{h\perp}$.
The axial four-spin vectors of the nucleon and the hadron, $S$ and $S_h$ respectively, satisfy $S^2=S_h^2=-1$ and $S\cd P=S_h\cd P_h=0$.
The differential cross section for SIDIS under one photon exchange can then be written as (see e.g.~\cite{Barone:2003fy})
\begin{align}
\frac{d^5\s}{dx_B\,dy\,dz_h\,d^2\vecb P_{h\perp}} &=
\frac{\pi \a_{em}^2}{2Q^4} y L_{\m\n} W^{\m\n}
\,.
\end{align}
The leptonic tensor $L_{\m\n}$ is
\begin{align}\label{eq:leptonictensor}
L_{\m\n} &= 
2\le(
k_\m k'_\n + k_\n k'_\m - g_{\m\n} k\cdot k'
\ri)
+ 2i\l_l \e_{\m\n\r\s}l^\r q^\s
\,,
\end{align}
where we have summed over the spin of the final lepton, $s_{l'}$.
The hadronic tensor $W^{\m\n}$ is given by
\begin{align}
W^{\m\n} &=
\frac{1}{(2\pi)^4}\frac{1}{z} \sum_X
\int \frac{d^3 P_X}{(2\pi)^3 2E_X}
(2\pi)^4 \d^{(4)}(P+q-P_h-P_X)
\sandwich{PS}{J^{\m\dagger}(0)}{X;P_hS_h}
\sandwich{X;P_hS_h}{J^\n(0)}{PS}
\nn\\
&=	
\frac{1}{z} \sum_{\,\,X}\!\!\!\!\!\!\!\!\int
\int \frac{d^4r}{(2\pi)^4} e^{iq\cdot r}
\sandwich{PS}{J^{\m\dagger}(r)}{X;P_hS_h}
\sandwich{X;P_hS_h}{J^\n(0)}{PS}
\,,
\end{align}
where the sum over the undetected hadrons in the final state, $X$, includes as well the integration over $P_X$.

The first step of factorization of the SIDIS hadronic tensor is done by matching the full QCD current~\footnote{We consider the case of a one photon exchange. The extension to $W$ and $Z$ bosons exchange is straightforward.}
\begin{align}
J_{\rm QCD}^\m = \sum_q e_q \bar \psi \g^\m \psi
\,,
\end{align}
onto the $q_T$-dependent one,
\begin{align}
\label{eq:SCETCURRENT}
J_{\rm SCET}^\m &=
C(Q^2/\m^2) \sum_q e_q
\bar \xi_\bn \tilde W_\bn^{T}  \tilde S_\bn^{T\dagger} \g^\m S_n^T \tilde W_n^{T\dagger}\xi_n
\,,
\end{align}
which contains soft and collinear modes.
The Wilson coefficient $C(Q^2/\m^2)$ can be extracted from the finite terms of the calculation of the (full QCD) quark form factor in pure dimensional regularization, and it is known up to ${\cal O}(\as^2)$~\cite{Idilbi:2006dg}.
See Appendix~\ref{app:hard} for more details.
For SIDIS kinematics the relevant Wilson lines, essential to insure gauge invariance among regular and singular gauges~\cite{Idilbi:2010im,GarciaEchevarria:2011md} are:
\begin{align}
&\tilde W_{n(\bn)}^T = \tilde T_{n(\bn)} \tilde W_{n(\bn)}\,,
\nn\\
&\tilde W_{n} (x) = \bar P \exp \left[-ig \int_{0}^{\infty} ds\, \nb \cdot A_n (x+\bn s)\right]\,,
\nn\\
&\tilde T_{n} (x) = \bar P \exp \left[-ig \int_{0}^{\infty} d\tau\, \vec l_\perp \cdot \vec A_{n\perp} (x^+,\infty^-,\vec x_\perp+\vec l_\perp \tau)\right]\,,
\nn\\
&\tilde T_{\bn} (x) = P \exp \left[-ig \int_{0}^{\infty} d\tau\, \vec l_\perp \cdot \vec A_{\bn\perp} (\infty^+,x^-,\vec x_\perp+\vec l_\perp \tau)\right]
\,,
\end{align}
and
\begin{align}
&S_{n}^T = T_{sn(s\bn)} S_{n}\,,
\quad\quad\quad\quad
\tilde S_{\bn}^T = \tilde T_{sn(s\bn)} \tilde S_{\bn}\,,
\nn\\
&S_n (x) = P \exp \left[i g \int_{-\infty}^0 ds\, n \cdot A_s (x+s n)\right]\,,
\nn \\
&T_{sn} (x) = P \exp \left[i g \int_{-\infty}^0 d\tau\, \vec l_\perp \cdot \vec A_{s\perp} (\infty^+,0^-,\vec x_\perp+\vec l_\perp \tau)\right]\,,
\nn \\
&T_{s\bn} (x) = P \exp \left[i g \int_{-\infty}^0 d\tau\, \vec l_\perp \cdot \vec A_{s\perp} (0^+,\infty^-,\vec x_\perp+\vec l_\perp \tau)\right]\,,
\nn\\
&\tilde S_\bn (x) = P\exp\le[-ig\int_{0}^{\infty} ds\, \bn \cdot A_s(x+\bn s) \ri]\,,
\nn\\
&\tilde T_{sn} (x) = P\exp\le[-ig\int_{0}^{\infty} d\t\, \vec l_\perp \cdot \vec A_{s\perp}(\infty^+,0^-,\vec x_\perp+\vec l_\perp\t) \ri]\,,
\nn\\
&\tilde T_{s\bn} (x) = P\exp\le[-ig\int_{0}^{\infty} d\t\, \vec l_\perp \cdot \vec A_{s\perp}(0^+,\infty^-,\vec x_\perp+\vec l_\perp\t) \ri]
\,.
\end{align}
$T_{sn(s\bn)}$ appears for the gauge choice $n \cdot A_s=0$ ($\bn \cdot A_s=0$), and the rest of the Wilson lines appearing in Eq.~(\ref{eq:SCETCURRENT}) are obtained by exchanging $n \leftrightarrow \bn$ and path-ordering $P$ with anti-path-ordering $\bP$.

One of the key ingredients of the  SCET machinery is the decoupling of the Hilbert space  of the partonic states into three subspaces corresponding to $n$-collinear, $\bn$-collinear and soft modes.
After this decoupling, standard manipulations lead to the following form of the hadronic tensor
\begin{align}
\label{eq:wfactor}
W^{\m\n} =
H(Q^2/\m^2) \frac{1}{N_c} \sum_q e_q 
\int \frac{d^4r}{(2\pi)^4} e^{iq\cdot r}\,
{\rm Tr}\big[
\Phi^{(0)}(r;P,S) \g^\m\,
\D^{(0)}(r;P_h,S_h) \g^\n\,
\big]
S(r)
+ {\cal O}\le(\frac{q_T}{Q}\ri)
\,,
\end{align}
where $H(Q^2/\m^2)=|C(Q^2/\m^2)|^2$ and
\begin{align}
\Phi_{ij}^{(0)}(r;P,S) &=
\le.
\sandwich{PS}{\le[\bar\x_{nj} \tilde W^T_n\ri](r)
\le[\tilde W_n^{T\dagger} \x_{ni}\ri](0)}{PS}
\ri|_{\rm zb\,\,subtracted}
\,,
\nn\\
\D_{ij}^{(0)}(r;P_h,S_h) &=
\frac{1}{z} \sum_{\,\,X}\!\!\!\!\!\!\!\!\int
\sandwich{0}{\le[\tilde W_\bn^{T\dagger} \x_{\bn i}\ri](r)}{X;P_hS_h}
\le.
\sandwich{X;P_hS_h}{\le[\bar\x_\bn \tilde W^T_{\bn j}\ri](0)}{0}
\ri|_{\rm zb\,\,subtracted}
\,,
\nn\\
S(r) &=\frac{1}{N_c}
\sandwich{0}{{\rm Tr}\, \le[S_n^{T\dagger} \tilde S_\bn^T \ri](r)\le[\tilde S^{T\dagger}_\bn S_n^T\ri](0)}{0}
\,.
\end{align}
The ``zb-subtracted'' stands for zero-bin subtraction which means that one needs to subtract the soft momentum modes (``zero-bin''  in SCET nomenclature) contributions from the naively calculated collinear matrix elements, thereby obtaining the so-called ``pure collinear'' matrix elements. 
Within SCET formalism, zero-bin subtractions were first introduced in Ref.~\cite{Manohar:2006nz}. 
In full QCD analysis, the issue of double counting was treated in Ref.~\cite{Collins:1999dz} through ``soft function subtraction'' (see also Ref.~\cite{Collins:2011zzd}). 
On the equivalence of the QCD and SCET treatments see Refs.~\cite{Lee:2006nr,Idilbi:2007ff,Idilbi:2007yi}.

In the region of large transverse momentum, $q_T\sim Q$, the factorized hadronic tensor in Eq.~\eqref{eq:wfactor} receives corrections through the so-called ``Y-term'' (see e.g. Section 13.12 in Ref~\cite{Collins:2011zzd}).
From now on we will omit this term and concentrate on the role of TMD functions and their evolution.

Since the incoming and outgoing quarks are $n$-collinear and $\bn$-collinear, respectively, the virtual photon momentum is hard, $q=k_\bn-k_n\sim Q(1,1,\l)$, and thus in the exponential in Eq.~(\ref{eq:wfactor}) we have $r\sim (1/Q)(1,1,1/\l)$.
Then, we need to Taylor expand the previous result and consider only the leading order contributions in $\lambda$.
Thus we get 
\begin{align}
W^{\m\n} &=
H(Q^2/\m^2) \frac{2}{N_c} \sum_q e_q 
\int d^2 \vecb k_{n\perp}d^2 \vecb k_{\bn\perp}d^2 \vecb k_{s\perp}
\d^{(2)}(\vecb q_\perp+\vecb k_{n\perp}-\vecb k_{\bn\perp}+\vecb k_{s\perp})
\,
\nn\\
&\times
{\rm Tr}\big[
\Phi^{(0)}(x,\vecb k_{n\perp},S) \g^\m\,
\D^{(0)}(z,\vecb k_{\bn\perp},S_h) \g^\n\,
\big]
S(\vecb k_{s\perp})
\,,
\end{align}
where
\begin{align}
\Phi_{ij}^{(0)}(x,\vecb k_{n\perp},S) &= 
\frac{1}{2} \int\frac{dy^-d^2\vecb y_\perp}{(2\pi)^3}
e^{-i(\frac{1}{2}y^-k_n^+-\vecbe y _\perp \cdot \vecbe k _{n\perp})}
\le.
\sandwich{PS}{\le[\bar\x_{nj} \tilde W^T_n\ri](0^+,y^-,\vecbp y)
\le[\tilde W_n^{T\dagger} \x_{ni}\ri](0)}{PS}
\ri|_{\rm zb\,\,subtracted}
\,,
\nn\\
\D_{ij}^{(0)}(z,\vecb{\hat P}_{h\perp},S_h) &=
\frac{1}{2} \int\frac{dy^+d^2\vecb y_\perp}{(2\pi)^3}
e^{i(\frac{1}{2}y^+k_\bn^--\vecbe y _\perp \cdot \vecbe k _{\bn\perp})}
\nn\\
&\times
\frac{1}{z} \sum_{\,\,X}\!\!\!\!\!\!\!\!\int
\sandwich{0}{\le[\tilde W_\bn^{T\dagger} \x_{\bn i}\ri](y^+,0^-,\vecb y_\perp)}{X;P_hS_h}
\le.
\sandwich{X;P_hS_h}{\le[\bar\x_\bn \tilde W^T_{\bn j}\ri](0)}{0}
\ri|_{\rm zb\,\,subtracted}
\,,
\nn\\
S(\vecb k_{s\perp}) &=
\int\frac{d^2\vecb y_\perp}{(2\pi)^2}
e^{i\vecbe y _\perp \cdot \vecbe k _{s\perp}}
\frac{1}{N_c}
\sandwich{0}{{\rm Tr}\, \le[S_n^{T\dagger} \tilde S_\bn^T \ri](0^+,0^-,\vecb y_\perp)\le[\tilde S^{T\dagger}_\bn S_n^T\ri](0)}{0}
\,.
\end{align}
For the $\Phi$ correlator we have $k_n^+=xP^+$, while for the $\D$ correlator we have $k_\bn^-={\hat P}_h^{-}/z$ and $\vecb k_{n\perp}=-\vecb {\hat P}_{h\perp}/z$.
$\vecb {\hat P}_{h\perp}$ can be interpreted as the transverse momentum of the outgoing hadron $h$ in a frame where the fragmenting quark has no transverse momentum.
On the other hand, $\vecb k_{\bn\perp}$ can be interpreted as the transverse momentum of the fragmenting quark in a frame where the outgoing hadron has no transverse momentum.
Thus one should notice the difference between $\vecb P_{h\perp}$, the transverse momentum of the hadron with respect to the photon, and $\vecb {\hat P}_{n\perp}$.

When calculated perturbatively (i.e. partonically) the three matrix elements above contain, individually, rapidity divergences.  
Those divergences are neither ultraviolet nor long-distance ones and, in principle, are not sensitive to confining dynamics. 
As argued in Ref.~\cite{GarciaEchevarria:2011rb}, such divergences appear in each one of the soft and collinear matrix elements contained in the factorization theorem, and they can be removed by articulating a particular combination of the soft and collinear matrix elements.

In order remove rapidity divergences from the sixteen TMDPDFs and TMDFFs, we split the soft function into two pieces~\cite{Echevarria:2012js},
\begin{align}\label{eq:splitting}
\tilde{S}(b_T;\frac{Q^2\mu^2}{\D^+\D^-},\m^2) &=
\tilde{S}_{-}\le(b_T;\z_F,\m^2;\D^-\ri)
\tilde{S}_{+}\le(b_T;\z_D,\m^2;\D^+\ri)
\,,
\nn\\
\tilde{S}_{-}\le(b_T;\z_F,\m^2;\D^-\ri) &=
\sqrt{\tilde S\left(\frac{\D^-}{p^+},\a\frac{\D^-}{\bar {p}^-}\right)}
\,,
\nn\\
\tilde{S}_{+}\le(b_T;\z_D,\m^2;\D^+\ri) &=
\sqrt{\tilde S\left(\frac{1}{\a}\frac{\D^+}{p^+},\frac{\D^+}{\bar {p}^-}\right)}
\,,
\end{align}
where in the soft functions under the square roots we have explicitly specified the dependence on the $\D$-regulator parameters that regulate the soft Wilson lines in the $n$- and $\bn$-directions.
More details on this splitting can be found in Sec.~\ref{sec:evolutiontmds}.
$\z_F$ and $\z_D$ are fractions of $Q^2$ satisfying $\z_F\z_D=Q^4$, where $\z_F=Q^2/\a$ and $\z_D=\a Q^2$ with $\a$ an arbitrary boost-invariant real number.
$p^+$ and $\bp^-$ stand for the two large collinear momentum components carried by the incoming and outgoing partons, respectively, that initiate the DIS hard reaction.
The superscript $\sim$ refers to quantities calculated in impact parameter space (IPS).

We emphasize the fact that the splitting of the soft function in rapidity space is a feature independent on any particular regulator~\cite{Echevarria:2012js}.
Although the arguments in that reference were based on a perturbative calculation performed with the $\D$-regulator, one could definitely use a different one to get to the same conclusion.

In order to properly define the TMDs, the two pieces of the soft function presented above are combined with the two quark correlators ($\Phi$ and $\D$). 
The resulting quantities are free from rapidity divergences and hence can be considered as a valid hadronic quantities.
Thus, the TMDPDFs are defined by
\begin{align}
\label{eq:TMDPDF}
F_{ij}(x,\vecb k_{n\perp},S;\z_F,\m^2;\D^-) &=
\int d^2\vecb b_\perp\, e^{i\vecbe b_\perp \cd \vecbe k_{n\perp}}\,
\tilde{\Phi}_{ij}^{(0)}(x,\vecbp b,S;\m^2;\D^-)\,
\tilde{S}_{-}(b_T;\z_F,\m^2;\D^-)
\,,
\end{align}
while for the TMDFFs we have
\begin{align}
\label{eq:TMDFF}
D_{ij}(z,\vecb{\hat P}_{h\perp},S_h;\z_D,\m^2;\D^+) &=
\int d^2\vecb b_\perp\, e^{-i\vecbe b_\perp \cd \vecbe k_{\bn\perp}}\,
\tilde{\D}_{ij}^{(0)}(z,\vecbp b,S_h;\m^2;\D^+)\,
\tilde{S}_{+}(b_T;\z_D,\m^2;\D^+)
\,.
\end{align}
With the definitions above we can write the hadronic tensor as
\begin{align}\label{eq:hadtensor}
W^{\m\n} &=
H(Q^2/\m^2) \frac{2}{N_c} \sum_q e_q 
\int d^2 \vecb k_{n\perp}d^2 \vecb k_{\bn\perp}
\d^{(2)}(\vecb q_\perp+\vecb k_{n\perp}-\vecb k_{\bn\perp})\,
\nn\\
&\times
{\rm Tr}\big[
F(x,\vecb k_{n\perp},S;\z_F,\m^2)\, 
\g^\m\,
D(z,\vecb{\hat P}_{h\perp},S_h;\z_D,\m^2)\,
\g^\n
\big]
\,.
\end{align}

Before continuing our analysis we would like to comment on the content of Eqs.~(\ref{eq:TMDPDF}-\ref{eq:TMDFF}). 
The above definition of the different TMDPDFs has been first introduced in Ref.~\cite{Echevarria:2012js}.
In this sense this is not a new result. 
However the above definition for TMDFFs can be considered as a generalization of the formalism of Refs.~\cite{GarciaEchevarria:2011rb,Echevarria:2012pw,Echevarria:2012js} to the case of unpolarized and the spin
 fragmentation functions.
In Appendix~\ref{app:tmdff_nlo} we present a next to leading order calculation of the unpolarized TMDFF and we show explicitly that it is free from rapidity divergences. 
We also perform, in Appendix~\ref{sec:tmdffope}, an OPE onto the integrated (or ``collinear'') fragmentation function (FF) and obtain the matching coefficient between the two for large transverse-momentum.

One could also extend the formulation given in Ref.~\cite{Collins:2011zzd} in order to properly define all the leading-twist TMDs.
The basics would be the same: split the soft function into two pieces and combine them with the collinear correlators to build well-defined quantities, free from rapidity divergences.

In Ref.~\cite{Boer:1997nt} a spin decompositions (for the different Dirac structure) was performed for the correlators $\Phi$ and $\D$. 
Such decompositions allow us to define sixteen TMD correlators at leading-twist: eight for initial state matrix elements and another eight for the analogous final state ones. 
Given the fact that the soft function introduced earlier is spin-independent, then the same spin-decompositions carry over straightforwardly for the well-defined TMDPDFs and TMDFFs ($F$ and $D$ respectively) in Eqs.~(\ref{eq:TMDPDF}-\ref{eq:TMDFF}), which contain the soft factor in them as explained above. 
Below we present the same spin decompositions as in Ref.~\cite{Boer:1997nt}, both in momentum space as in IPS, however it should be understood that we are referring, throughout the rest of this work, to the newly defined objects.
This distinction is crucial, since the properties of the two referred objects are completely different, as their QCD evolution is.
The inclusion of the soft function in the definition of the TMDs is a must in order to obtain a well-defined hadronic quantities.

Given the above and with the notation $F^{[\G]}\equiv\frac{1}{2}\rm{Tr}\,\le(F\,\G\ri)$, where $\Gamma$ is a generic combination of Dirac matrices, one has the following decomposition for the TMDPDFs: 
\begin{align}
F^{[\g^+]}(x,\vecb k_{n\perp},S;\z_F,\m^2) &= 
f_1(x,k_{nT}^2;\z_F,\m^2)
- \frac{\epsilon_\perp^{ij}\vecb k_{n\perp i} \vecb S_{\perp j}}{M_N}\,
f_{1T}^\perp(x,k_{nT}^2;\z_F,\m^2)
\,,
\nn\\
F^{[\g^+\g_5]}(x,\vecb k_{n\perp},S;\z_F,\m^2) &= 
\lambda\, g_{1L}(x,k_{nT}^2;\z_F,\m^2)
+ \frac{(\vecb k_{n\perp}\cdot\vecb S_\perp)}{M_N}\,
g_{1T}(x,k_{nT}^2;\z_F,\m^2)
\,,
\nn\\
F^{[i\s^{i+}\g_5]}(x,\vecb k_{n\perp},S;\z_F,\m^2)  &= 
\vecb S_\perp^i\,h_{1}(x,k_{nT}^2;\z_F,\m^2)
+ \frac{\lambda\, \vecb k_{n\perp}^i}{M_N}\,
h_{1L}^\perp(x,k_{nT}^2;\z_F,\m^2)
\nn\\ 
&
- \frac{\left(\vecb k_{n\perp}^i \vecb k_{n\perp}^j + \frac{1}{2}\vecb k_{n\perp}^2g_\perp^{ij}\right)
\vecb S_{\perp j}}{M_N^2}\,
h_{1T}^\perp(x,k_{nT}^2;\z_F,\m^2)
- \frac{\epsilon_\perp^{ij} \vecb k_{n\perp j}}{M_N}\,
h_1^\perp(x,k_{nT}^2;\z_F,\m^2)
\,.
\end{align}
Analogously (and using $D^{[\G]}=\frac{1}{2}\rm{Tr}\,\le(D\,\G\ri)$) we have the following decomposition for TMDFFs:
\begin{align}\label{eq:tmdffdecompositionk}
D^{[\g^-]}(z,\vecb{\hat P}_{h\perp},S_h;\z_D,\m^2) &= 
D_1(z,{\hat P}_{hT}^2;\z_D,\m^2)
- \frac{\epsilon_\perp^{ij}\vecb k_{\bn\perp i} \vecb S_{h\perp j}}{M_h}\,D_{1T}^\perp(z,{\hat P}_{hT}^2;\z_D,\m^2)
\,,
\nn\\
D^{[\g^-\g_5]}(z,\vecb{\hat P}_{h\perp},S_h;\z_D,\m^2) &= 
\l\,G_{1L}(z,{\hat P}_{hT}^2;\z_D,\m^2)
+ \frac{(\vecb k_{\bn\perp}\cdot\vecb S_{h\perp})}{M_h}\,
G_{1T}(z,{\hat P}_{hT}^2;\z_D,\m^2)
\,,
\nn\\
D^{[i\s^{i-}\g_5]}(z,\vecb{\hat P}_{h\perp},S_h;\z_D,\m^2)  &= 
\vecb S_{h\perp}^i\,
H_{1}(z,{\hat P}_{hT}^2;\z_D,\m^2)
+ \frac{\lambda\, \vecb k_{\bn\perp}^i}{M_h}\,
H_{1L}^\perp(z,{\hat P}_{hT}^2;\z_D,\m^2)
\nn\\ 
&
- \frac{\left(\vecb k_{\bn\perp}^i \vecb k_{\bn\perp}^j + \frac{1}{2}\vecb k_{\bn\perp}^2g_\perp^{ij}\right)
\vecb S_{h\perp j}}{M_h^2}\,
H_{1T}^\perp(z,{\hat P}_{hT}^2;\z_D,\m^2)
- \frac{\epsilon_\perp^{ij} \vecb k_{\bn\perp j}}{M_h}\,
H_1^\perp(z,{\hat P}_{hT}^2)
\,.
\end{align}

Let us now express the hadronic tensor in terms of $\vecb P_{n\perp}$, which is the transverse momentum of the hadron with respect to the photon direction.
The transverse momentum of the virtual photon, $\vecb q_\perp$, is related to $\vecb P_{h\perp}$ by $\vecb q_\perp \approx -\vecb P_{h\perp}/z$ up to ${\cal O}(1/Q^2)$ corrections.
Using this relation and $\vecb k_{\bn\perp}=-\vecb{\hat P}_{h\perp}/z$, we can write
\begin{align}\label{eq:hadtensor2}
W^{\m\n} &=
H(Q^2/\m^2) \frac{2}{N_c} \sum_f e_f 
\int d^2 \vecb k_{n\perp} \frac{d^2\vecb{\hat P}_{h\perp}}{z^2}\,
\d^{(2)}\le(\frac{\vecb P_{h\perp}}{z}-\vecb k_{n\perp}-\frac{\vecb{\hat P}_{h\perp}}{z}\ri)\,
\nn\\
&\times
{\rm Tr}\big[
F_{f/N}(x,\vecb k_{n\perp},S;\z_F,\m^2)\, 
\g^\m\,
D_{h/f}(z,\vecb{\hat P}_{h\perp},S_h;\z_D,\m^2)\,
\g^\n
\big]
\,.
\end{align}
The trace in Eq.~(\ref{eq:hadtensor2}) can be decomposed into different Dirac structures by means of  Fierz transformations 
\begin{eqnarray}
\lefteqn{4(\gamma^\mu)_{jk}(\gamma^\nu)_{li}=}\nonumber\\
&&\left[ {\bf 1}_{ji}{\bf 1}_{lk}
+(i\gamma_5)_{ji}(i\gamma_5)_{lk}-(\gamma^\alpha)_{ji}(\gamma_\alpha)_{lk}
-(\gamma^\alpha\gamma_5)_{ji}(\gamma_\alpha\gamma_5)_{lk}
+\case{1}{2} (i\sigma_{\alpha\beta}\gamma_5)_{ji}(i\sigma^{\alpha\beta}
\gamma_5)_{lk}\right]g^{\mu\nu} \nonumber\\
&&+(\gamma^{\{ \mu})_{ji}(\gamma^{\nu\} })_{lk}+
(\gamma^{\{ \mu}\gamma_5)_{ji}(\gamma^{\nu\} }\gamma_5)_{lk}+
(i\sigma^{\alpha\{\mu}\gamma_5)_{ji}(i{\sigma^{\nu\}} }_\alpha \gamma_5)_{lk}
+\ldots,\label{Dirac}
\end{eqnarray}
where we have kept only the terms symmetric under the exchange of
$\m$ and $\n$.
If one considers the scattering of a nucleon by an unpolarized lepton, the leptonic tensor in eq.~\eqref{eq:leptonictensor} is symmetric, and thus we only need the symmetric part of the hadronic tensor.
With this decomposition we get
\begin{align}\label{eq:hadtensortmds}
W^{\m\n} &=
H(Q^2/\m^2) \frac{2}{N_c} \sum_f e_f 
\int d^2\vecb k_{n\perp} \frac{d^2\vecb {\hat P}_{h\perp}}{z^2}
\d^{(2)}\le(\frac{\vecb P_{h\perp}}{z}-\vecb k_{n\perp}-\frac{\vecb{\hat P}_{h\perp}}{z}\ri)\,
\nn\\
&\times
\le[
\le(
F_{f/N}^{[\g^+]}(x,\vecb k_{n\perp},S)\,
D_{h/f}^{[\g^-]}(z,\vecb{\hat P}_{h\perp},S_h) 
+ F_{f/N}^{[\g^+\g_5]}(x,\vecb k_{n\perp},S)\, 
D_{h/f}^{[\g^-\g_5]}(z,\vecb{\hat P}_{h\perp},S_h)
\ri) g_\perp^{\m\n}
\ri.
\nn\\
&\le.
+
F_{f/N}^{[i\s^{i+}\g_5]}(x,\vecb k_{n\perp},S)\,
D_{h/f}^{[i\s^{j-}\g5]}(z,\vecb{\hat P}_{h\perp},S_h)
\le(g_{\perp i}^{\{\m}\, g^{\n\}}_j - g_{\perp ij}\, g_\perp^{\m\n}\ri)
\ri]
\,,
\end{align}
where $g_\perp^{\m\n}=g^{\m\n}-\frac{1}{2}(n^\m\bn^\n+n^\n\bn^\m)$ and we have used the properties of $n$-collinear and $\bn$-collinear fields: $\nslash\x_n=0$, $\bnslash\x_\bn=0$, $\nslash\x_\bn=\x_\bn$ and $\bnslash\x_n=\x_n$.

In IPS, where convolutions become simple products, the hadronic tensor is expressed as follows:
\begin{align}\label{eq:hadtensortmds2}
W^{\m\n} &=
H(Q^2/\m^2) \frac{2}{N_c} \sum_f e_f 
\int \frac{d^2\vecb b}{(2\pi)^2}\, 
e^{-i\vecbe b\cd\vecbe P_{h\perp}/z}
\nn\\
&\times
\le[
\le(
\tilde{F}_{f/N}^{[\g^+]}(x,\vecb b_\perp,S;\z_F,\m^2)\,
\tilde{D}_{h/f}^{[\g^-]}(z,\vecb b_\perp,S_h;\z_D,\m^2) 
+ 
\tilde{F}_{f/N}^{[\g^+\g_5]}(x,\vecb b_\perp,S;\z_F,\m^2)\, 
\tilde{D}_{h/f}^{[\g^-\g_5]}(z,\vecb b_\perp,S_h;\z_D,\m^2)
\ri) g_\perp^{\m\n}
\ri.
\nn\\
&\le.
+
\tilde{F}_{f/N}^{[i\s^{i+}\g_5]}(x,\vecb b_\perp,S;\z_F,\m^2)\,
\tilde{D}_{h/f}^{[i\s^{j-}\g5]}(z,\vecb b_\perp,S_h;\z_D,\m^2)
\le(g_{\perp i}^{\{\m}\, g^{\n\}}_j - g_{\perp ij}\, g_\perp^{\m\n}\ri)
\ri]
\,,
\end{align}
where
\begin{align}\label{eq:fouriertransforms}
\tilde{F}_{f/N}(x,\vecb b_\perp,S;\z_F,\m^2) &=
\int d^2\vecb k_{n\perp}\,
e^{i\vecbe b_\perp\cd\vecbe k_{n\perp}}
F_{f/N}(x,\vecb k_{n\perp},S;\z_F,\m^2)
\,,
\nn\\
\tilde{D}_{h/f}(z,\vecb b_\perp,S_h;\z_D,\m^2) &=
\frac{1}{z^2} \int d^2\vecb{\hat P}_{h\perp}\, 
e^{i\vecbe b_\perp\cd\vecbe{\hat P}_{h\perp}/z}
D_{h/f}(z,\vecb{\hat P}_{h\perp},S_h;\z_D,\m^2)
\,.
\end{align}
Since the evolution of all TMDs will be discussed in IPS, then it is useful to introduce the TMDPDFs and TMDFFs in that space. 
When Fourier transforming to IPS we get (similar expressions can be found in Ref.~\cite{Boer:2011xd})
\begin{align}\label{eq:tmdpdfdecompositionips}
\tilde{F}_{f/N}^{[\g^+]}(x,\vecb b_{\perp},S;\z_F,\m^2) &= 
\tilde{f}_1(x,b_{T}^2;\z_F,\m^2)
- \frac{\epsilon_\perp^{ij}\vecb b_{\perp i} \vecb S_{\perp j}}{ib_T M_N}\,
\tilde{f}_{1T}^{\perp(1)}(x,b_{T}^2;\z_F,\m^2)
\,,
\nn\\
\tilde{F}_{f/N}^{[\g^+\g_5]}(x,\vecb b_{\perp},S;\z_F,\m^2) &= 
\lambda\, \tilde{g}_{1L}(x,b_{T}^2;\z_F,\m^2)
+ \frac{(\vecb b_{\perp}\cdot\vecb S_\perp)}{ib_T M_N}\,
\tilde{g}_{1T}^{(1)}(x,b_{T}^2;\z_F,\m^2)
\,,
\nn\\
\tilde{F}_{f/N}^{[i\s^{i+}\g_5]}(x,\vecb b_{\perp},S;\z_F,\m^2) &= 
\vecb S_\perp^i\,\tilde{h}_{1}(x,b_{T}^2;\z_F,\m^2)
+ \frac{\lambda\, \vecb b_{\perp}^i}{ib_T M_N}\,
\tilde{h}_{1L}^{\perp(1)}(x,b_{T}^2;\z_F,\m^2)
\nn\\ 
&
- \frac{\left(\vecb b_{\perp}^i \vecb b_{\perp}^j + \frac{1}{2}b_{T}^2g_\perp^{ij}\right)
\vecb S_{\perp j}}{(i)^2 b_T^2 M_N^2}\,
\tilde{h}_{1T}^{\perp(2)}(x,b_{T}^2;\z_F,\m^2)
- \frac{\epsilon_\perp^{ij} \vecb b_{\perp j}}{ib_T M_N}\,
\tilde{h}_1^{\perp(1)}(x,b_{T}^2;\z_F,\m^2)
\,,
\end{align}
and
\begin{align}\label{eq:tmdffdecompositionips}
\tilde{D}_{h/f}^{[\g^-]}(z,\vecb b_{\perp},S_h;\z_D,\m^2) &= 
\tilde{D}_1(z,b_{T}^2;\z_D,\m^2)
- \frac{\epsilon_\perp^{ij} \vecb b_{\perp i} \vecb S_{h\perp j}}{(-ib_T)M_h}\, \tilde{D}_{1T}^{\perp(1)}(z,b_{T}^2;\z_D,\m^2)
\,,
\nn\\
\tilde{D}_{h/f}^{[\g^-\g_5]}(z,\vecb b_{\perp},S_h;\z_D,\m^2) &= 
\l\,\tilde{G}_{1L}(z,b_{T}^2;\z_D,\m^2)
+ \frac{(\vecb b_{\perp}\cd\vecb S_{h\perp})}{(-ib_T)M_h}\,
\tilde{G}_{1T}^{(1)}(z,b_{T}^2;\z_D,\m^2)
\,,
\nn\\
\tilde{D}_{h/f}^{[i\s^{i-}\g_5]}(z,\vecb b_{\perp},S_h;\z_D,\m^2) &= 
\vecb S_{h\perp}^i\,
\tilde{H}_{1}(z,b_{T}^2;\z_D,\m^2)
+ \frac{\lambda\, \vecb b_{\perp}^i}{(-ib_T)M_h}\,
\tilde{H}_{1L}^{\perp(1)}(z,b_{T}^2;\z_D,\m^2)
\nn\\ 
&
- \frac{\left(\vecb b_{\perp}^i \vecb b_{\perp}^j + \frac{1}{2} b_{T}^2g_\perp^{ij}\right)
\vecb S_{h\perp j}}{(-ib_T)^2 M_h^2}\,
\tilde{H}_{1T}^{\perp(2)}(z,b_{T}^2;\z_D,\m^2)
- \frac{\epsilon_\perp^{ij} \vecb b_{\perp j}}{(-ib_T)M_h}\,
\tilde{H}_1^{\perp(1)}(z,b_{T}^2;\z_D,\m^2)
\,,
\end{align}
where the superscript $(n)$ stands for the $n$-th derivative with respect to $b_T$.
Thus, consistent with Eq.~\eqref{eq:fouriertransforms}, for any of the eight functions in the decomposition of $\tilde{F}_{f/N}$ we have
\begin{align}
\tilde{F}^{(n)}(b_T^2) &=
\le(\frac{\pd}{\pd b_T}\ri)^{n}
\tilde{F}(b_{T}^2)
=
\le(\frac{\pd}{\pd b_T}\ri)^{n}
\int d^2\vecb k_{n\perp}\,
e^{i\vecbe b_\perp\cd\vecbe k_{n\perp}}\, F(k_{nT}^2)
\,,
\end{align}
while for any of the eight functions in the decomposition of $\tilde{D}_{h/f}$ we have
\begin{align}
\tilde{D}^{(n)}(b_T^2) &=
\le(\frac{\pd}{\pd b_T}\ri)^{n}
\tilde{D}(b_{T}^2)
=
\le(\frac{\pd}{\pd b_T}\ri)^{n}
\frac{1}{z^2} \int d^2\vecb {\hat P}_{n\perp}\,
e^{i\vecbe b_\perp\cd\vecbe {\hat P}_{h\perp}/z}\, D({\hat P}_{hT}^2)
\,.
\end{align}
Recall that for the fragmentation function we have $\vecb k_{\bn\perp}=-\vecb{\hat P}_{h\perp}/z$, and we have used this relation to get Eq.~\eqref{eq:tmdffdecompositionips} from Eq.~\eqref{eq:tmdffdecompositionk}.

Depending on whether we are interested in spin-averaged quantities or spin asymmetries for either the incoming or outgoing hadrons, and depending on the directions of those spins (longitudinal and/or transverse), different terms of the expansions of $F_{f/N}$ and $D_{h/f}$ will appear in the hadronic tensor.
However the hard part , which is just a multiplicative factor of  the soft and the two collinear contributions and is a polynomial in $\log(Q^2/\mu^2)$,    \ is the same among all possible pairings of TMDPDF-TMDFF.

\section{Evolution of TMDPDFs and TMDFFs}
\label{sec:evolutiontmds}

The scale evolution of the different TMDs is governed through their anomalous dimensions which are defined as follows
\begin{align}
\label{eq:evo1}
\frac{d}{d\ln\m}\ln\tilde{F}^{[\G]}_{f/N}(x,\vecb b_\perp,S;\z_F,\m^2) &\equiv
\g_F\le(\as(\m),\ln\frac{\z_F}{\m^2}\ri)
\,,
\nn\\
\frac{d}{d\ln\m}\ln\tilde{D}^{[\G]}_{h/f}(z,\vecb b_\perp,S_h;\z_D,\m^2) &\equiv
\g_D\le(\as(\m),\ln\frac{\z_D}{\m^2}\ri)
\,.
\end{align}
Based on the factorized hadronic tensor in IPS given in Eq.~\eqref{eq:hadtensortmds2}, the evolution of the TMDs with respect to the factorization scale $\m$ is related to that of the hard part. Since the hadronic tensor does not depend on the factorization scale, the anomalous dimensions $\g_F$ and $\g_D$ are related to the one of the hard part, $\g_H$, through
\begin{align}
\label{eq:evo2}
\g_H &= \frac{d}{d\ln\m} H(Q^2/\m^2) = 
2\G_{\rm cusp}(\as(\m))\ln\frac{Q^2}{\m^2} + 2\g^V(\as(\m))
\,,
\nn\\
&=
- \g_F\le(\as(\m),\ln\frac{\z_F}{\m^2}\ri) 
- \g_D\le(\as(\m),\ln\frac{\z_D}{\m^2}\ri)
\,,
\end{align}
and thus
\begin{align}
\label{eq:evo3}
\g_F\le(\as(\m),\ln\frac{\z_F}{\m^2}\ri) &=
-\G_{\rm cusp}(\as(\m))\ln\frac{\z_F}{\m^2} - \g^V(\as(\m))
\,,
\nn\\
\g_D\le(\as(\m),\ln\frac{\z_D}{\m^2}\ri) &=
-\G_{\rm cusp}(\as(\m))\ln\frac{\z_D}{\m^2} - \g^V(\as(\m))
\,.
\end{align}
It should be mentioned that the splitting of $\gamma_H$ into $\gamma_F$ and $\gamma_D$ given in the last equation is unique following the restriction of $\z_F\z_D=Q^4$. 
The coefficients of the perturbative expansions of $\G_{\rm cusp}$ and $\g^V$ are known up to three loops and they are collected in~\cite{Echevarria:2012pw}.

On the other hand, the TMDs depend as well on $Q^2$ through the variables $\z_F$ and $\z_D$. 
This can be easily verified, e.g., by considering the NLO results for the unpolarized TMDPDF (see Eq.~(21) in Ref.~\cite{Echevarria:2012js}) or for the unpolarized TMDFF (see Eq.~\eqref{eq:tmdfffirstorderb} in Appendix~\ref{app:tmdff_nlo}). 
We next discuss the evolution of all TMDs with respect to $Q^2$, or equivalently $\z_F$ and $\z_D$.

The starting point is Eqs.~(\ref{eq:TMDPDF})-(\ref{eq:TMDFF}). In IPS where the convolution becomes a simple product, one has the following:
\begin{align}
\label{eq:TMDPDF2}
\ln F_{ij}(x,\vecb b_{\perp},S;\z_F,\m^2;\D^-) &=
\ln \tilde{\Phi}_{ij}^{(0)}(x,\vecbp b,S;\m^2;\D^-)+\ln \tilde{S}_{-}(b_T;\z_F,\m^2;\D^-)
\end{align}
and

\begin{align}
\label{eq:TMDFF2}
\ln D_{ij}(z,\vecb{b}_{\perp},S_h;\z_D,\m^2;\D^+) &=
\ln \tilde{\D}_{ij}^{(0)}(z,\vecbp b,S_h;\m^2;\D^+)+
\ln\tilde{S}_{+}(b_T;\z_D,\m^2;\D^+)
\,.
\end{align}
We notice that the $\zeta$-dependence in Eqs.~({\ref{eq:TMDPDF2}) and~(\ref{eq:TMDFF2}) lies completely in the soft factors, while the pure collinear contributions ($\tilde{\Phi}^{(0)}$ and $\tilde{\Delta}^{(0)}$) are free from any $\zeta$-dependence. This observation is important. Each pure collinear contribution depends solely on one collinear sector: $n$-collinear for the TMDPDFs and $\bn$-collinear for the TMDFFs\footnote{This is not the case for the naive collinear contributions, since such quantities involve soft contamination in each of them. This soft contamination ``connects'' the two collinear sectors and thus a non-valid $Q^2$-dependence appears in the collinear contributions to both the TMDPDFs and the TMDFFs. Thus avoiding double counting is crucial.}. 
As such, it is impossible to generate any $Q^2$-dependence in those quantities since the only way that the $Q^2$ can appear (either in the collinear or the soft factors) is through the (boost invariant) combination of $p^+\bp^-=Q^2$ (here we are assuming that we are in the Breit frame). 
On the other hand the soft gluon radiation has no preferred collinear direction (both light-cone momentum components have the same scaling) and the soft factors do include $Q^2$-dependence through a term of the form $\log(\D^+\D^-/Q^2\mu^2)$ (see Eq.~(18) in Ref.~\cite{Echevarria:2012js}). 
Moreover, in Ref.~\cite{Echevarria:2012js}, where we considered the DY kinematics, it was shown that to all orders in perturbation theory, $\ln\tilde{S}$ has a single logarithmic dependence on $\ln(\D^+\D^-/Q^2\mu^2)$, 
\begin{align}\label{eq:softsplitting}
\ln\tilde{S} &=
{\cal R}_s(b_T,\as) + 2D(b_T,\as)\,\ln\left(\frac{\D^+\D^-}{Q^2\mu^2}\right)
\,.
\end{align}
Thus this function can be split into
\begin{align}
\label{eq:TM3}
\ln\tilde{S}_-=
\frac{1}{2}{\cal R}_s(b_T,\alpha_s) 
+ D(b_T,\alpha_s)\ln\left(\frac{(\D^-)^2}{\z_F\mu^2}\right)
\,,
\end{align}
and
\begin{align}
\label{eq:FF3}
\ln\tilde{S}_+=
\frac{1}{2}{\cal R}_s(b_T,\alpha_s) 
+ D(b_T,\alpha_s)\ln\left(\frac{(\D^+)^2}{\z_D\mu^2}\right)
\,,
\end{align}
where, as already mentioned before, $\z_F=Q^2/\a$ and $\z_D=\a Q^2$ with $\a$ an arbitrary boost-invariant real number.

Given the fact that the soft function is Hermitian and its logarithm has single logarithm of $Q^2$ to all orders in perturbation theory, then when going from time-like (DY) kinematics to space-like ones (DIS), it is evident that the soft function is universal. 
Thus the arguments of Ref.~\cite{Echevarria:2012js} for the splitting of the soft function carry over straightforwardly to SIDIS kinematics. Moreover the $D$-term is also universal among the DIS and DY kinematics.
Combining this observation with Eqs.~(\ref{eq:TMDPDF2}) and~(\ref{eq:TMDFF2}), we get that the $Q^2$-dependence of the TMDPDFs and TMDFFs is governed by:
\begin{align}\label{eq:zetaevolution}
\frac{d}{d\ln\z_F}\ln\tilde{F}^{[\G]}_{f/N}(x,\vecb b_\perp,S;\z_F,\alpha_s) &=
- D(b_T;\alpha_s)
\,,
\nn\\
\frac{d}{d\ln\z_D}\ln\tilde{D}^{[\G]}_{h/f}(z,\vecb b_\perp,S_h;\z_D,\alpha_s) &=
- D(b_T;\alpha_s)
\,.
\end{align}
It can be easily verified that, given Eqs.~(\ref{eq:evo1}-\ref{eq:evo2}-\ref{eq:evo3}) and the $\m$-independence of the hadronic tensor we have
\begin{align}\label{eq:Devolution}
\frac{dD}{d\ln\m} &= \G_{\rm cusp}
\,.
\end{align}

Since the soft function is spin-independent and universal~\footnote{Previously, Collins and Metz got to the same conclusion in Ref.~\cite{Collins:2004nx} while employing a different line of reasoning.}, and given the perturbative arguments above, by extrapolation from small to large values of $b_T$ we arrive to the conclusion that the evolution of all TMDs with respect to $Q^2$ is governed by a single universal and spin-independent quantity, namely the $D$-term (and $\g_H$). 
This is one of the main results of this work.
Next we discuss the $D$-term.

As it is clear from the above discussion regarding small vs. large values of $b_T$, the $D$-term contains perturbative and non-perturbative information.
Given Eq.~(\ref{eq:softsplitting}), the coefficients of the perturbative expansion of the $D$-term can be completely determined by performing a perturbative calculation of the partonic soft function.
In Ref.~\cite{GarciaEchevarria:2011rb} we explained how to obtain the NLO coefficient of the D-term, which is necessary to obtain the evolution kernel up to next-to-next-to logarithmic accuracy (NNLL), from a fixed order calculation of the (full QCD) DY cross-section.
However, as it was shown in Ref.~\cite{Echevarria:2012pw}, even after resumming the large logarithms in the perturbative expansion of the $D$-term,  the resummed $D$ has a finite range of convergence in IPS.
Thus one needs to parameterize (or ``model'') this quantity for large values of $b_T$. 
However, since, as argued above, the soft function is universal and spin-independent, the considered non-perturbative model can be applied to parameterize the large $b_T$ region of the $D$-term, and hence the evolution kernel, regardless which TMD function we are considering.
This is also generally assumed within the standard Collins-Soper-Sterman approach~\cite{Collins:1984kg}, where the non-perturbative model for the Collins-Soper kernel is taken to be universal (see also Ref.~\cite{Aidala:2014hva}).

By setting a hard cutoff $b_{Tc}$ we can separate the two contributions to the $D$-term, and thus it can be written as
\begin{align}
D(b_T;\m) &=
D^{R}(b_T;\m)\,\theta(b_{Tc}-b_T) + D^{NP}(b_T)\,\theta(b_T-b_{Tc})
\,.
\end{align}
In Ref.~\cite{Echevarria:2012pw} it was found, while exploiting all the available perturbative information, the region in the IPS where the resummed $D$ ($D^R$) converges.
On the other hand, the values of the parameters that enter into the model for $D^{NP}$ should be extracted from fits to experimental data.

Regardless how the non-perturbative contribution to the $D$-term is parameterized, we can perform the evolution of all leading-twist TMDPDFs and TMDFFs consistently up to NNLL:
\begin{align}\label{eq:evolutiontmds}
\tilde{F}^{[\G]}_{f/N}(x,\vecb b_\perp,S;\z_{F,f},\m_f^2) &=
\tilde{F}^{[\G]}_{f/N}(x,\vecb b_\perp,S;\z_{F,i},\m_i^2)\,
\tilde R\le(b_T;\z_{F,i},\m_i^2,\z_{F,f},\m_f^2\ri)
\,,
\nn\\
\tilde{D}^{[\G]}_{h/f}(z,\vecb b_\perp,S_h;\z_{D,f},\m_f^2) &=
\tilde{D}^{[\G]}_{h/f}(z,\vecb b_\perp,S_h;\z_{D,i},\m_i^2)\,
\tilde R\le(b_T;\z_{D,i},\m_i^2,\z_{D,f},\m_f^2\ri)
\,,
\end{align}
where the evolution kernel $\tilde{R}$ is given by
\begin{align}\label{eq:evolkernel}
\tilde R\big(b;\z_i,\m_i^2,\z_f,\m_f^2\big) &=
\exp\le\{
\int_{\m_i}^{\m_f} \frac{d\bar\m}{\bar\m}\, 
\g\le(\as(\bar\m),\ln\frac{\z_f}{\bar\m^2} \ri)
\ri\}
\le( \frac{\z_f}{\z_i} \ri)^{-D\le(b_T;\m_i\ri)}
\,,
 \end{align}
with $(\z=\z_{F},\g=\g_F)$ for the TMDPDFs and $(\z=\z_{D},\g=\g_D)$ for the TMDFFs.
Notice that Eq.~\eqref{eq:evolutiontmds} is valid for all the sixteen functions  appearing in Eqs.~\eqref{eq:tmdpdfdecompositionips} and~\eqref{eq:tmdffdecompositionips}. 

When trying to implement the evolution kernel for different experiments, one needs to relate $\z_F$ and $\z_D$ to the physical scale $Q^2$. 
As already mentioned, for a given $Q^2$ we have the relation $\z_F\z_D=Q^4$.
Thus, whether we are considering anyone of the eight TMDPDFs or the eight TMDFFs, the $\zeta$ parameter has different values for a given $Q^2$. 
For all practical purposes, one considers a hadronic tensor where only the combination $\z_F\z_D$ appears in the product of two TMDs. 
Then, one can safely relate $\z_F$ and $\z_D$ to $Q^2$ by setting: $\z_F=\z_D=Q^2$, and when doing so, we replace the parameter $\zeta$  in Eq.~(\ref{eq:evolkernel}) with $Q^2$, which is the actual physical scale set by experiment. 
With this choice we can safely claim that all the sixteen TMDs have the same evolution kernel given in Eq.~(\ref{eq:evolkernel}).

In order to illustrate the application of Eq.~\eqref{eq:evolutiontmds}, let us consider Collins function $H_1^{\perp}$. 
One has to notice that in the decomposition of $\tilde{D}_{h/f}^{[\G]}$ in Eq.~\eqref{eq:tmdffdecompositionips} what appears is the first derivative of Collins function with respect to the impact parameter, and thus, consistently with Eq.~\eqref{eq:evolutiontmds}, we have that
\begin{align}\label{eq:dcollinsevolved}
\tilde{H}_1^{\perp(1)}(z,b_{T}^2;Q_f) &=
\tilde{H}_1^{\perp(1)}(z,b_{T}^2;Q_i)\,
\tilde{R}(b_T;Q_i,Q_f)
\,,
\end{align}
where for simplicity we have set $\m^2=\z=Q^2$.
It is this derivative and not the function itself which would appear in the factorized hadronic tensor in IPS in Eq.~\eqref{eq:hadtensortmds2}, and thus it is the derivative of Collins function which is evolved simply by multiplying it with the evolution kernel $\tilde{R}$.
Given the Fourier transforms in Eq.~\eqref{eq:fouriertransforms}, the derivative of Collins function in IPS is related to the function in momentum space through 
\begin{align}\label{eq:dcollinsips}
\tilde{H}_1^{\perp(1)}(z,b_T^2;Q) &=
\frac{-2\pi}{z^3}
\int_0^\infty d{\hat P}_{hT}\, 
{\hat P}_{hT}^2\,
J_1(b_T{\hat P}_{hT}/z)\,
H_1^{\perp}(z,{\hat P}_{hT}^2;Q)
\,,
\end{align}
while the inverse relation is given by
\begin{align}\label{eq:collinsmomentum}
H_1^{\perp}(z,{\hat P}_{hT}^2;Q) &=
\frac{-z}{2\pi{\hat P}_{hT}}
\int_0^\infty d{b}_{T}\, 
b_{T}\,
J_1(b_T{\hat P}_{hT}/z)\,
\tilde{H}_1^{\perp(1)}(z,b_T^2;Q)
\,.
\end{align}
Thus, if we have a parameterization for $H_1^{\perp}(z,{\hat P}_{hT}^2;Q_i)$ at some initial scale, we should first calculate its derivative $\tilde{H}_1^{\perp(1)}(z,b_T^2;Q_i)$, then evolve it by multiplying it with the evolution kernel and finally obtain $H_1^{\perp}(z,{\hat P}_{hT}^2;Q_f)$ at a higher scale $Q_f$.

\begin{figure}[t]
\begin{center}
\includegraphics[width=0.45\textwidth]{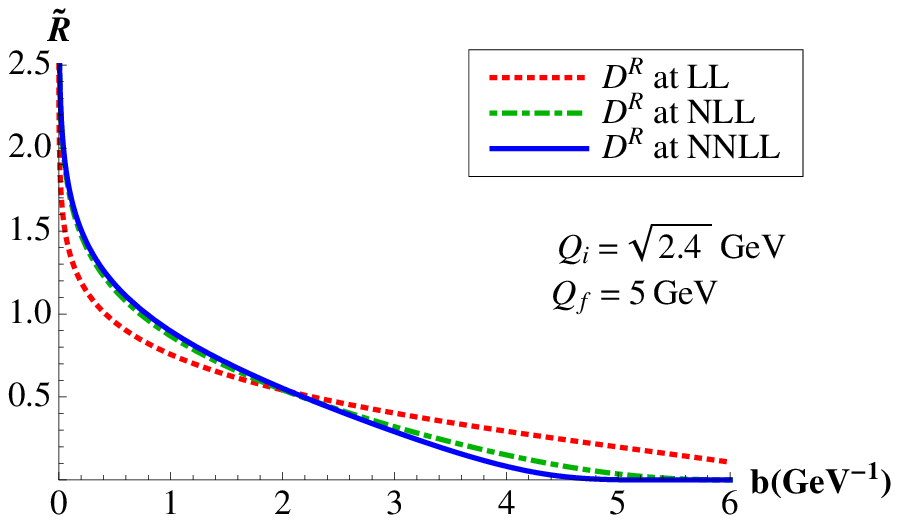}
\quad\quad\quad
\includegraphics[width=0.45\textwidth]{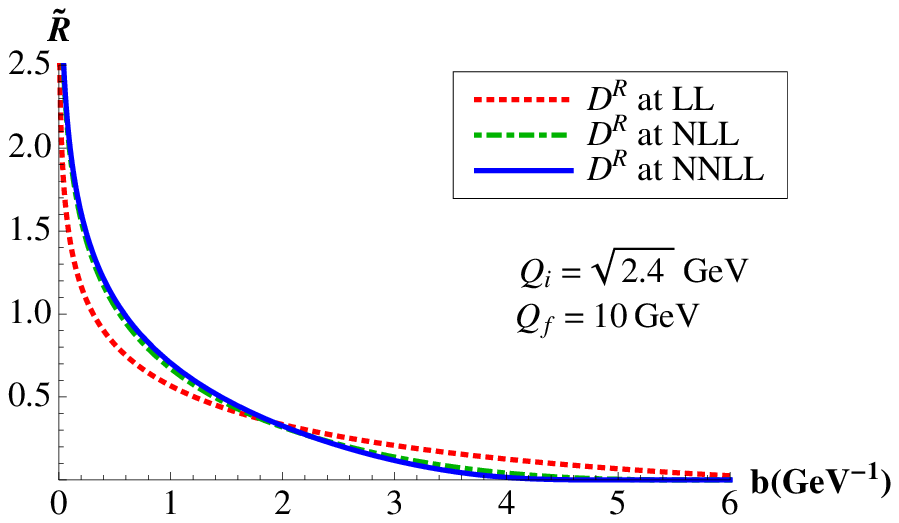}
\\
\hspace{0cm}(a)\hspace{9cm}(b)\hspace{6cm}
\end{center}
\caption{\it
Evolution kernel $\tilde{R}$ for $Q_i=\sqrt{2.4}~{\rm GeV}$ and $Q_f=[5;10]~{\rm GeV}$ with the resummed $D$, $D^R$, introduced in Ref.~\cite{Echevarria:2012pw}.
}
\label{fig:kernel}
\end{figure}

After obtaining the evolution kernel in Eq.~\eqref{eq:evolkernel} and once we have managed to separate the perturbative and non-perturbative contributions to the $D$-term, it is important to notice that the TMDs themselves also contain perturbative information when the transverse momentum is large ($k_T \gg \lqcd$).
In other words, one can perform an operator product expansion (OPE) of the TMDs onto collinear functions and thus extract the dependence on the transverse momentum in terms of a perturbatively calculable Wilson coefficient:
\begin{align}\label{eq:tmdsope}
\tilde{T}(b_T;\z,\m^2) &=
\tilde{C}(b_T;\z;\m^2) \otimes t(\m^2)
+ {\cal O}\le(b_T\lqcd\ri)
\,.
\end{align}
The convolution refers to variables $x$ or $z$ for TMDPDFs or TMDFFs, respectively.
In the equation above we have schematically represented the OPE, where $\tilde{T}(b_T;\z,\m^2)$ stands for any one  of the sixteen functions presented in Eqs.~\eqref{eq:tmdpdfdecompositionips} and~\eqref{eq:tmdffdecompositionips}, and $t(\m^2)$ for the corresponding collinear function.
For instance, we could consider the unpolarized TMDPDF and match it onto the unpolarized collinear PDF (see e.g. Refs.~\cite{Aybat:2011zv,GarciaEchevarria:2011rb}); or the derivative of Sivers function and match it onto twist-3 collinear function (see e.g. Ref.~\cite{Aybat:2011ge}); or the TMD helicity and transversity functions and match them onto their collinear counterparts (see Ref.~\cite{Bacchetta:2013pqa}).
Given Eq.~\eqref{eq:softsplitting}, this general OPE can be further expanded in order to exponentiate the $\z$-dependence in the matching coefficient:
\begin{align}\label{eq:tmdsope2}
\tilde{T}(b_T;\z,\m^2) &=
\le(\frac{\z b_T^2}{4e^{-2\g_E}}\ri)^{-D(b_T;\m)}
\tilde{C}^{\Qslash}(b_T;\m^2) \otimes t(\m^2)
+ {\cal O}\le(b_T\lqcd\ri)
\,,
\end{align}
where $\tilde{C}^{\Qslash}$ stands for the part of the Wilson coefficient in Eq.~\eqref{eq:tmdsope} after the exponentiation of the $\z$-dependence.
We emphasize the fact that the OPE above holds only in the perturbative region of small $b_T$.
Thus one should impose a cutoff over $b_T$ and add a parameterization for the large $b_T$ region, which should be extracted from fitting to experimental data.

Now, if we combine the evolution kernel given in Eq.~\eqref{eq:evolkernel} with the OPE in Eq.~\eqref{eq:tmdsope2}, we can finally write the TMDs while expanding, explicitly, their perturbative content to the maximal extent:
\begin{align}
\tilde{T}(b_T;\z,\m^2) &=
\le[
\tilde{C}^{\Qslash}(b_T;\m_I^2)
\otimes t(\m_I^2)\ri]
\,
\exp\le\{
\int_{\m_I}^{\m} \frac{d\bar\m}{\bar\m}\, 
\g\le(\as(\bar\m),\ln\frac{\z}{\bar\m^2} \ri)
\ri\}
\le(\frac{\z b_T^2}{4e^{-2\g_E}}\ri)^{-D(b_T;\m_I)}
\,.
\end{align}
In order to minimize the effect of large logarithms in the perturbative parts, the best choice for the dummy scale $\m_I$ is $\m_I\sim k_T\sim 1/b_T$.
On the other hand, notice that this expression, as it stands, gives us the TMD $\tilde{T}$ in the region of small $b_T$.
Thus, in order to recover the complete range of the impact parameter we should include some cutoff over $b_T$, and at the same time be able to separate the perturbative and non-perturbative contributions to the $D$-term to the maximal extent, while resumming large logarithms to the highest possible logarithmic accuracy according to a well-defined resummation scheme.
Finally, one should add as well a model to account for the large $b_T$ region where the OPE in Eq.~\eqref{eq:tmdsope2} breaks down.

\section{Application: Evolution of Collins TMDFF}
\label{sec:collins}

\begin{figure}[t]
\begin{center}
\includegraphics[width=0.45\textwidth]{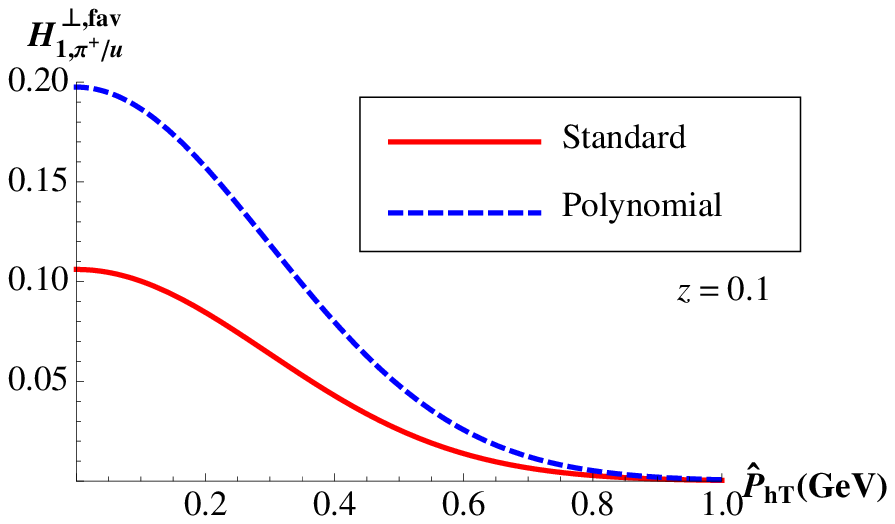}
\quad\quad\quad
\includegraphics[width=0.45\textwidth]{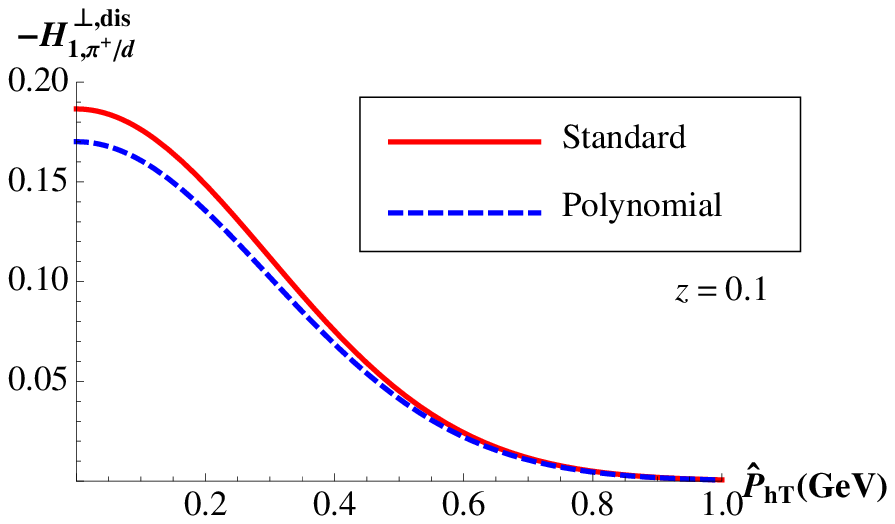}
\\
(a)\hspace{9cm}(b)
\end{center}
\caption{\it
Input models for Collins function at $Q_i=\sqrt{2.4}~{\rm GeV}$, favored case (a) and disfavored case (b).}
\label{fig:collins}
\end{figure}

\begin{figure}[t]
\begin{center}
\includegraphics[width=0.45\textwidth]{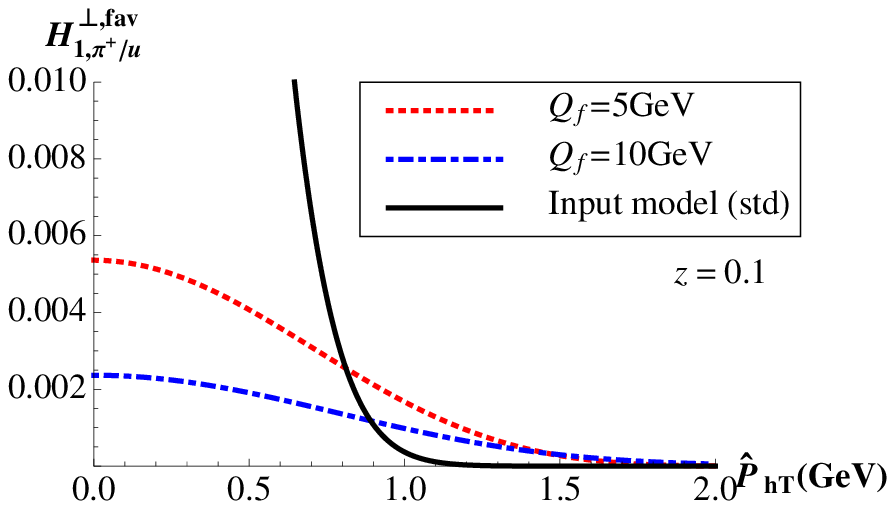}
\quad\quad\quad
\includegraphics[width=0.45\textwidth]{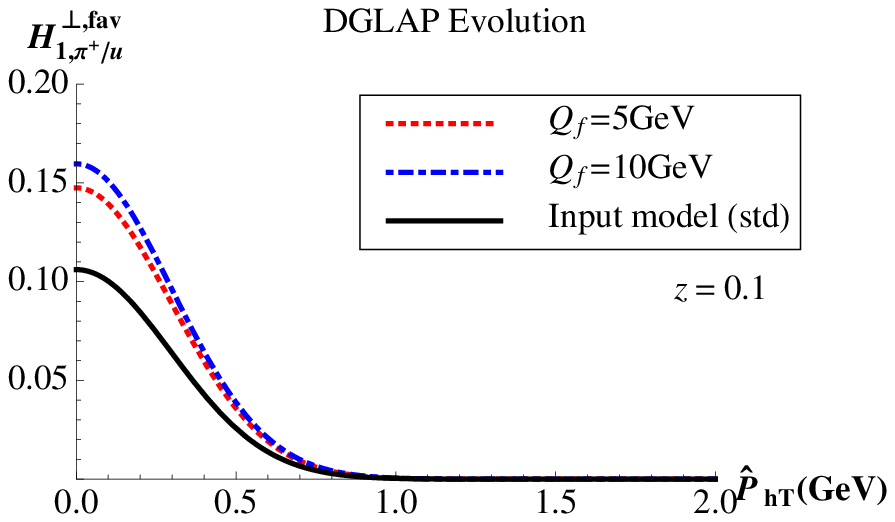}
\\
(a)\hspace{9cm}(b)
\\
\includegraphics[width=0.45\textwidth]{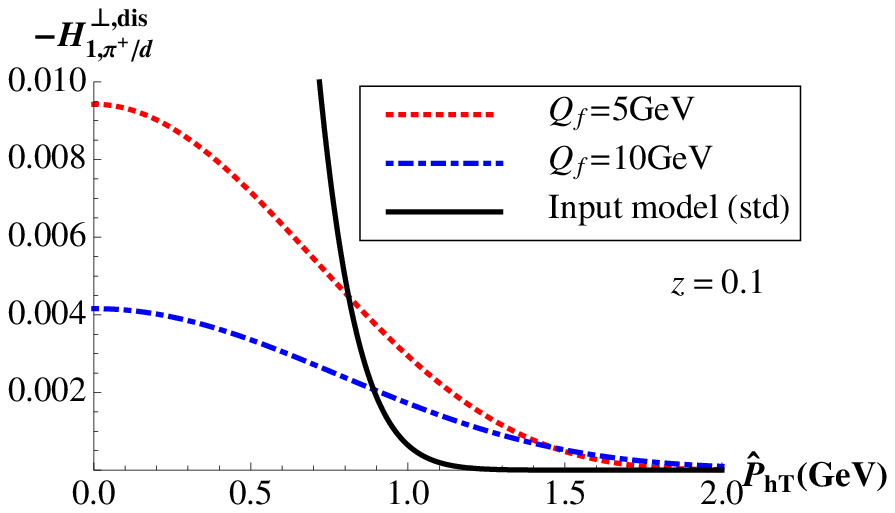}
\quad\quad\quad
\includegraphics[width=0.45\textwidth]{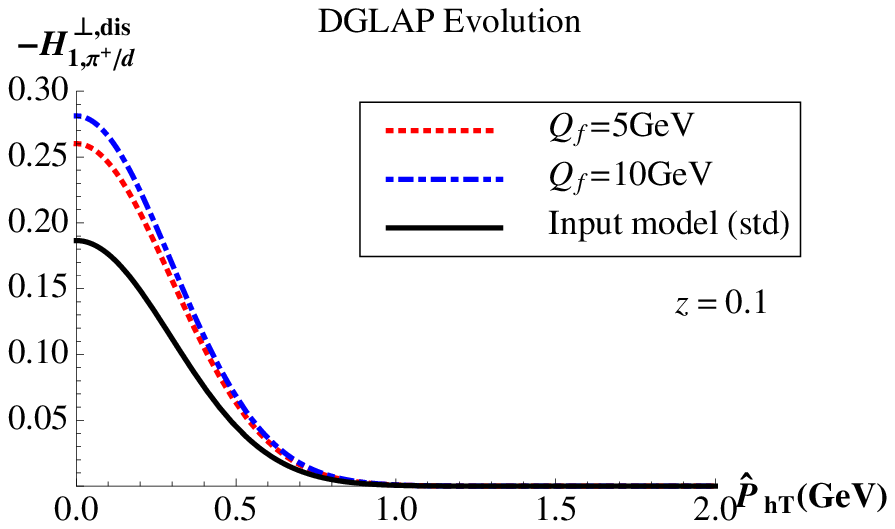}
\\
(c)\hspace{9cm}(d)
\end{center}
\caption{\it
Evolution at NNLL accuracy of the Collins function from $Q_i=\sqrt{2.4}~{\rm GeV}$ up to two different final scales.
(a) Standard input model in the favored case with proper QCD evolution for TMDs. 
(b) Standard input model in the favored case with DGLAP evolution for the collinear FF.
(c) Standard input model in the disfavored case with proper QCD evolution for TMDs. 
(d) Standard input model in the disfavored case with DGLAP evolution for the collinear FF.
}
\label{fig:collinsSTD_comparison}
\end{figure}

\begin{figure}[t]
\begin{center}
\includegraphics[width=0.45\textwidth]{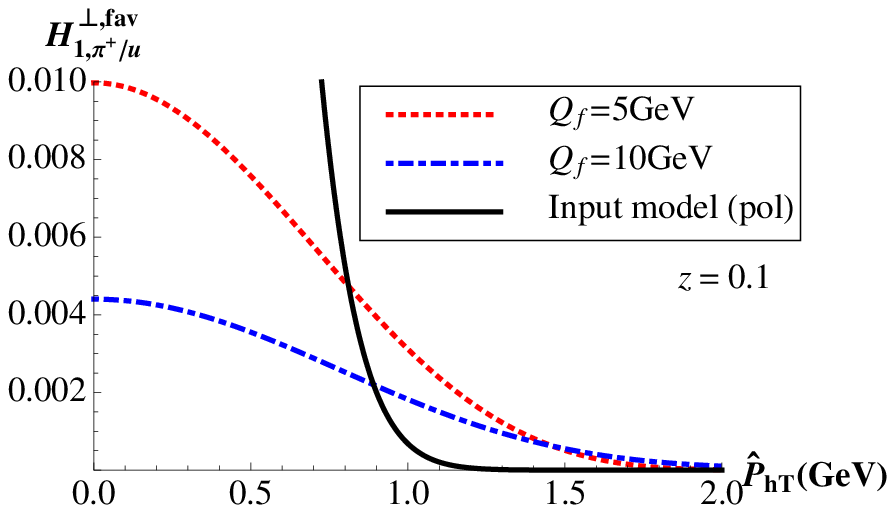}
\quad\quad\quad
\includegraphics[width=0.45\textwidth]{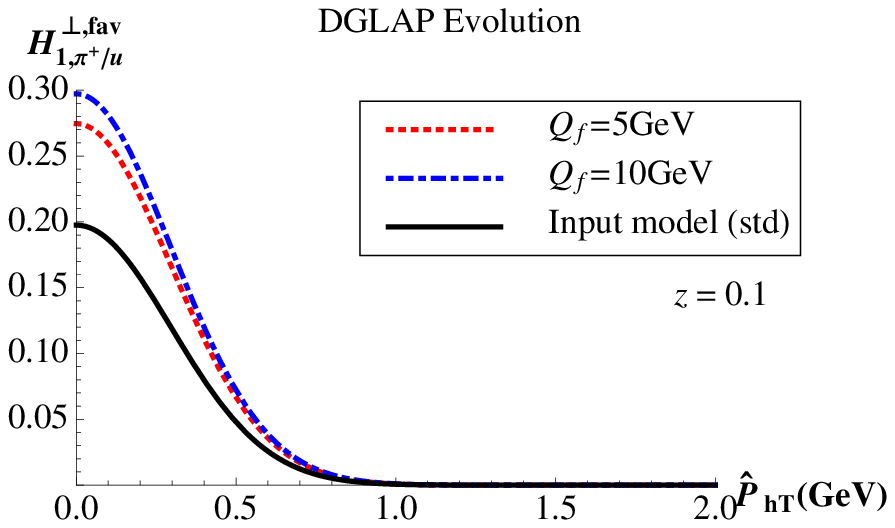}
\\
(a)\hspace{9cm}(b)
\\
\includegraphics[width=0.45\textwidth]{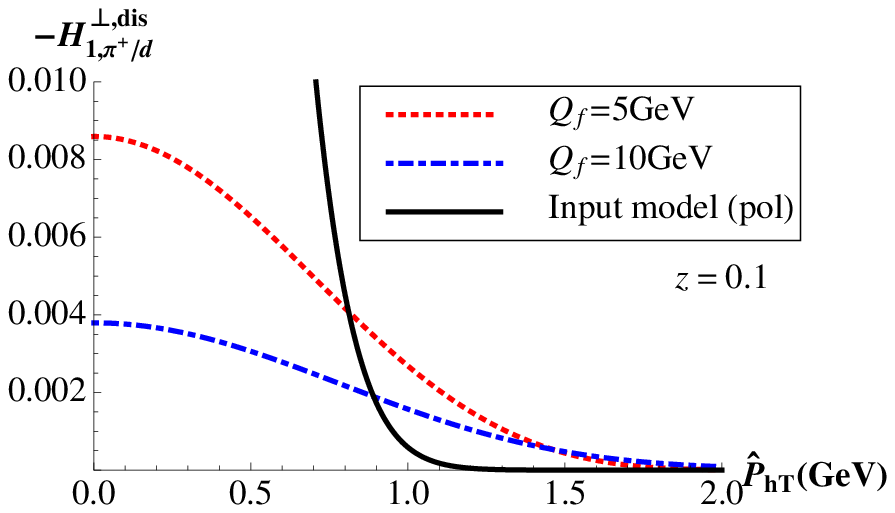}
\quad\quad\quad
\includegraphics[width=0.45\textwidth]{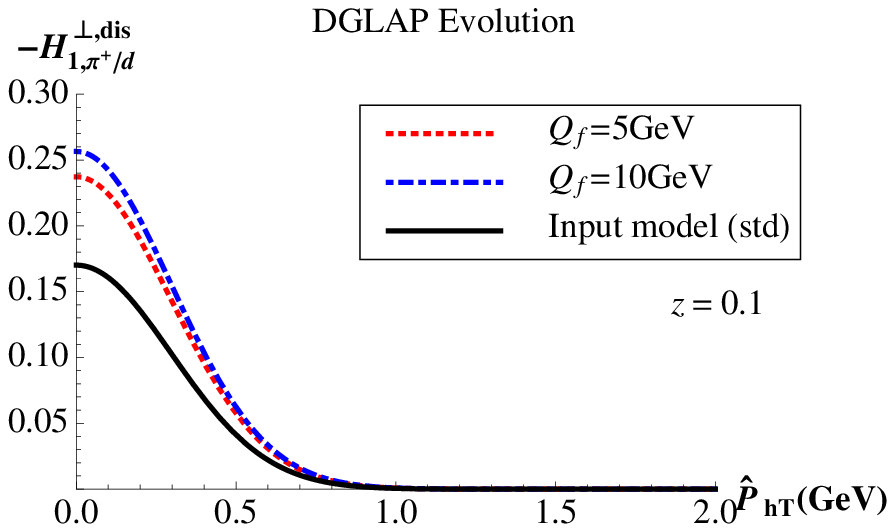}
\\
(c)\hspace{9cm}(d)
\end{center}
\caption{\it
Evolution at NNLL accuracy of the Collins function from $Q_i=\sqrt{2.4}~{\rm GeV}$ up to two different final scales.
(a) Polynomial input model in the favored case with proper QCD evolution for TMDs. 
(b) Polynomial input model in the favored case with DGLAP evolution for the collinear FF.
(c) Polynomial input model in the disfavored case with proper QCD evolution for TMDs. 
(d) Polynomial input model in the disfavored case with DGLAP evolution for the collinear FF.
}
\label{fig:collinsPOL_comparison}
\end{figure}

After explicitly deriving the evolution for all leading-twist (un-)polarized TMDPDFs and TMDFFs, which turns out to be driven by the same evolution kernel, our goal in this section is to illustrate its application by considering a particular polarized TMD function: Collins function.
In the literature one can find several examples of phenomenological studies of TMDs where, in order to deal with experimental data obtained at different scales, the approach taken is to evolve the collinear functions (PDF or FF) that enter into the parameterizations of the considered TMDs (see e.g.~\cite{Anselmino:2013rya,Anselmino:2013lza,Anselmino:2013vqa}).
In other words, the evolution is implemented through the standard DGLAP evolution kernels.
However, as we discuss below, the application of the proper QCD evolution gives very different results as compared with the implementation of DGLAP kernel.
Moreover, we apply the evolution consistently at NNLL accuracy, the highest possible one given the present knowledge we have of the perturbative ingredients that enter in the evolution kernels.

Notice that here we are referring to the ``modified'' definition of Collins function, consistent with Eq.~\eqref{eq:TMDFF} (also with Ref.~\cite{Collins:2011zzd}), since the one introduced in Ref.~\cite{Collins:1992kk} did not contain the proper soft factor.
Obviously the evolution of those two quantities is quite different, and below we consider the one defined as in Eq.~\eqref{eq:TMDFF}.

The authors in Ref.~\cite{Anselmino:2013vqa} present the last extraction of Collins function available in the literature.
They perform a global fit of data of azimuthal asymmetries, considering SIDIS, from HERMES and COMPASS collaborations, and electron-positron annihilation, from Belle Collaboration, in order to extract the Collins function and the transversity.
The data from the different collaborations are given at widely separated scales, and thus the implementation of the evolution of the relevant hadronic matrix elements becomes inevitable if one wants to interpret them properly.
As already mentioned, the authors apply the DGLAP evolution to the collinear functions, but we will anyway take the parameterization of Collins function they extract as our input at the lower scale and apply to it the proper TMD evolution.
In  future studies, it will be beneficial to revise previous phenomenological analyses while taking into account this evolution.

Before we actually proceed with the application of the evolution to Collins function obtained in Ref.~\cite{Anselmino:2013vqa}, we need to be careful with the different convention used in that work to define the Fourier transforms.
Instead of the convolution appearing in Eq.~\eqref{eq:hadtensor2}, they actually use $\d^{(2)}(\vecb P_{h\perp}-z\vecb k_{n\perp}-\vecb{\hat P}_{h\perp})$, and thus the consistent Fourier transforms are
\begin{align}\label{eq:fouriertransformsNEW}
\tilde{F}_{f/N}(x,z\vecb b_\perp,S;\z_F,\m^2) &=
\int d^2\vecb k_{n\perp}\,
e^{iz\vecbe b_\perp\cd\vecbe k_{n\perp}}
F_{f/N}(x,\vecb k_{n\perp},S;\z_F,\m^2)
\,,
\nn\\
\tilde{D}_{h/f}(z,\vecb b_\perp,S_h;\z_D,\m^2) &=
\int d^2\vecb{\hat P}_{h\perp}\, 
e^{i\vecbe b_\perp\cd\vecbe{\hat P}_{h\perp}}
D_{h/f}(z,\vecb{\hat P}_{h\perp},S_h;\z_D,\m^2)
\,.
\end{align}
Notice the difference with respect to Eq.~\eqref{eq:fouriertransforms}.
Then, the corresponding relations between the derivative of Collins function in IPS and the function itself in momentum space are
\begin{align}\label{eq:dcollinsipsNEW}
\tilde{H}_1^{\perp(1)}(z,b_T^2;Q) &=
-2\pi \int_0^\infty d{\hat P}_{hT}\, 
{\hat P}_{hT}^2\,
J_1(b_T{\hat P}_{hT})\,
H_1^{\perp}(z,{\hat P}_{hT}^2;Q)
\,,
\end{align}
and
\begin{align}\label{eq:collinsmomentumNEW}
H_1^{\perp}(z,{\hat P}_{hT}^2;Q) &=
\frac{-1}{2\pi{\hat P}_{hT}}
\int_0^\infty d{b}_{T}\, 
b_{T}\,
J_1(b_T{\hat P}_{hT})\,
\tilde{H}_1^{\perp(1)}(z,b_T^2;Q)
\,.
\end{align}
Those are the relations we use below in order to illustrate the effect of the QCD evolution on Collins function, taking as an input the model extracted in Ref.~\cite{Anselmino:2013vqa}.

Following the Trento convention~\cite{Bacchetta:2004jz},  Collins function is given by
\begin{equation}
\Delta^N D_{h/q^{\uparrow}}(z,{\hat P}_{hT}^2) =
\frac{2 {\hat P}_{hT}}{z M_h} \,  H_1^{\perp}(z,{\hat P}_{hT}^2)
\,,
\end{equation} 
where $H_1^{\perp}$ is the function that appears in the decomposition in Eq.~(\ref{eq:tmdffdecompositionips}). This function is parameterized in Ref.~\cite{Anselmino:2013vqa} as
\begin{align}
\Delta^N \! D_{h/q^\uparrow}(z,{\hat P}_{hT}^2) &= 
2 \, {\cal N}^{C}_{q}(z)\,
D_{h/q}(z)\, 
h({\hat P}_{hT})\,
\frac{e^{-{\hat P}_{hT}^2/{\avpht}}}{\pi\avpht}
\,,\quad\quad\quad
h({\hat P}_{hT}) =
\sqrt{2e}\,\frac{{\hat P}_{hT}}{M_{h}}\,
e^{-{{\hat P}_{hT}^2}/{M_{h}^2}}
\,.
\end{align}
Here $D_{h/q}(z)$ represents the collinear FF.
Two different parameterizations for ${\cal N}^{C}_q(z)$ are considered, called ``standard'' and ``polynomial'':
\begin{align}
{\cal N}^{C,std}_q(z) &= 
N^{C}_q \, z^{\gamma} (1-z)^{\delta} \,
\frac{(\gamma + \delta)^{(\gamma +\delta)}}
{\gamma^{\gamma} \delta^{\delta}}
\,,
\nn\\
{\cal N}^{C,pol}_q(z) &= 
N^{C}_q z \le[
(1-a-b) + az + bz^2
\ri]
\,.
\end{align}
As explained in Ref.~\cite{Anselmino:2013vqa} it is convenient for  fitting purposes to introduce ``favored'' and ``disfavored'' fragmentation functions, and thus Collins function $H_1^\perp$ is modeled as:
\begin{align}\label{eq:collins4models}
H_{1,\pi^+/u,\bar{d}}^{\perp,fav}(z,{\hat P}_{hT}^2;Q_i^2) &=
z\, {\cal N}^{C,fav}_{q}(z)\,
D_{\pi^+/u,\bar{d}}(z;Q_i^2)\, 
\sqrt{2e}\,
e^{-{{\hat P}_{hT}^2}/{M_{h}^2}}\,
\frac{e^{-{\hat P}_{hT}^2/{\avpht}}}{\pi\avpht}
\,,
\nn\\
H_{1,\pi^-/d,\bar{u}}^{\perp,fav}(z,{\hat P}_{hT}^2;Q_i^2) &=
z\, {\cal N}^{C,fav}_{q}(z)\,
D_{\pi^-/d,\bar{u}}(z;Q_i^2)\, 
\sqrt{2e}\,
e^{-{{\hat P}_{hT}^2}/{M_{h}^2}}\,
\frac{e^{-{\hat P}_{hT}^2/{\avpht}}}{\pi\avpht}
\,,
\nn\\
H_{1,\pi^+/d,\bar{u}}^{\perp,dis}(z,{\hat P}_{hT}^2;Q_i^2) &=
z\, {\cal N}^{C,dis}_{q}(z)\,
D_{\pi^+/d,\bar{u}}(z;Q_i^2)\, 
\sqrt{2e}\,
e^{-{{\hat P}_{hT}^2}/{M_{h}^2}}\,
\frac{e^{-{\hat P}_{hT}^2/{\avpht}}}{\pi\avpht}
\,,
\nn\\
H_{1,\pi^-/u,\bar{d}}^{\perp,dis}(z,{\hat P}_{hT}^2;Q_i^2) &=
z\, {\cal N}^{C,dis}_{q}(z)\,
D_{\pi^-/u,\bar{d}}(z;Q_i^2)\, 
\sqrt{2e}\,
e^{-{{\hat P}_{hT}^2}/{M_{h}^2}}\,
\frac{e^{-{\hat P}_{hT}^2/{\avpht}}}{\pi\avpht}
\,.
\end{align}
In Tables~\ref{tab:parameters_std} and~\ref{tab:parameters_pol} one can find the parameters that appear in the models above, and in Fig.~\ref{fig:collins} we show the correspondent input Collins functions at the initial scale $Q_i=\sqrt{2.4}~{\rm GeV}$.
The evolved Collins function in momentum space is obtained by using  Eqs.~\eqref{eq:dcollinsevolved}, \eqref{eq:dcollinsips} and~\eqref{eq:collinsmomentum}.

\begin{table}[h]
\begin{center}
\begin{tabular}{ccccc}
\hline
$N_q^{C,fav}=0.49$ &\quad\quad
$N_q^{C,dis}=-1.00$ &\quad\quad
$\g=1.06$ &\quad\quad
$\d=0.07$ &\quad\quad
$M_h^2=1.50$
\\
\hline
\end{tabular}
\end{center}
\caption{Best fit parameters from~\cite{Anselmino:2013vqa} for the standard parameterization of ${\cal N}_q^{C,std}(z)$.
We do not specify the uncertainties since we do not use them.}
\label{tab:parameters_std}
\end{table}

\begin{table}[h]
\begin{center}
\begin{tabular}{ccccc}
\hline
$N_q^{C,fav}=1.00$ &\quad\quad
$N_q^{C,dis}=-1.00$ &\quad\quad
$a=-2.36$ &\quad\quad
$b=2.12$ &\quad\quad
$M_h^2=0.67$
\\
\hline
\end{tabular}
\end{center}
\caption{Best fit parameters from~\cite{Anselmino:2013vqa} for the polynomial parameterization of ${\cal N}_q^{C,pol}(z)$.
We do not specify the uncertainties since we do not use them.}
\label{tab:parameters_pol}
\end{table}

In Figs.~\ref{fig:collinsSTD_comparison} and~\ref{fig:collinsPOL_comparison} we compare the evolved Collins function by applying the proper QCD evolution for TMDs, on one hand, and the DGLAP evolution on the other.
We consider both standard and polynomial parameterizations in the favored and disfavored cases, as it appeares in Eq.~\eqref{eq:collins4models}.
The QCD evolution is applied without the implementation of any non-perturbative model for the evolution kernel, since, as explained in Ref.~\cite{Echevarria:2012pw} and shown in Fig.~\ref{fig:kernel}, its effect is negligible if the considered initial and final scales are well separated, as it is our case.
As can be easily noticed, there is a substantial difference between the QCD evolution and DGLAP.
While the QCD evolution induces a fast decrease of the function and broadens its width (see as well Refs.~\cite{Aybat:2011zv,Aybat:2011ge}), DGLAP evolution induces an enhancement.
This was also observed in~\cite{Anselmino:2012aa} while considering the evolution of Sivers function.

In conclusion, and in order to properly interpret experimental data and extract from them sensible results for the TMDs, one should apply the correct evolution which, given the results shown in this work, it is now available for all leading-twist TMDPDFs and TMDFFs.
On the other hand, it is also worth emphasizing that the evolution of TMDs, or in other words, the resummation of large logarithms, should be applied consistently within a resummation scheme~\cite{Echevarria:2012pw}. 
Not only the anomalous dimensions, $\g_{F,D}$ and $\G_{\rm cusp}$, and the $D$ terms should be expanded accordingly, taking care of the difference between the cusp and non-cusp terms in $\g_{F,D}$, but the matching coefficients $H$ and $\tilde{C}$ as well.
For instance, for a resummation up to NLL accuracy, one should take the matching coefficients $H$ and $\tilde{C}$ at tree-level, the anomalous dimension $\g^V$ and the $D$ term at 1-loop and the cusp anomalous dimension $\G_{\rm cusp}$ at two-loops.

\section{Conclusions}
\label{sec:conclusions}

Using the formalism of effective field theories we have derived a factorization theorem for SIDIS process.
The relevant soft function, which is shown to be universal between DY and SIDIS kinematics, is split in rapidity space into two pieces and each one is then combined with one of the two collinear sectors. 
This combination allows us to obtain a well-defined TMDPDFs and TMDFFs in the sense that all rapidity divergences cancel.  
We have argued, while extensively discussing the properties of the soft function, that the last statement is valid for the sixteen relevant TMDs considered in this work. 
In particular, we have shown this fact by explicitly calculating the unpolarized TMDFF at NLO, and its matching onto the collinear FF. We emphasize that this successful matching could not have been attained without the contribution of the relevant soft contribution to the collinear one.

By considering the properties of the pure collinear and soft matrix elements, we have shown that the evolution kernel for all the leading-twist TMDs is identical.
The current knowledge of the perturbative ingredients that enter into this kernel allows us to perform the evolution of all TMDs while resumming  large logarithms consistently up to NNLL accuracy.
We have illustrated the application of this kernel by considering one particular polarized TMDFF: Collins function and pointed out the difference between applying the proper TMD evolution compared with the widely used DGLAP one. The differences among the two are clear from our results and this is one of the main results of this work.

By probing hadrons at different scales and processes (SIDIS, DY or electron-positron annihilation) we can unravel their inner momentum and spin structure, which is encoded by those TMDs.
This research is actively being pursued by HERMES (DESY), COMPASS (CERN), CLAS (JLAB), Belle (KEK) or BaBar (SLAC) collaborations, among others.
The LHC and the future electron-ion collider can also be of very much help in understanding the internal structure of hadrons.
All the previously mentioned experiments run at different energies and probe hadrons at different scales, thus, in order to properly interpret experimental data, it is absolutely necessary to know and implement the evolution of the TMDs involved in a given process.
Concluding, this work opens the door for revising previous phenomenological analyzes of spin asymmetries and performing new ones, while considering the proper QCD evolution of such quantities.

\section*{Acknowledgements}
We thank A.~Vladimirov for useful comments.
I.~S. is supported by the Spanish MEC, FPA2011-27853-CO2-02.
M.~G.~E. is supported by the ``Stichting voor Fundamenteel Onderzoek der Materie (FOM)'', which is financially supported by the ``Nederlandse Organisatie voor Wetenschappelijk Onderzoek (NWO)''.
A.~I. is supported by the US Department of Energy under grant number DE-SC0008745.

\appendix



 
\section{TMDFF at NLO}
\label{app:tmdff_nlo}

\begin{figure*}[t]
\begin{center}
\includegraphics[width=\textwidth]{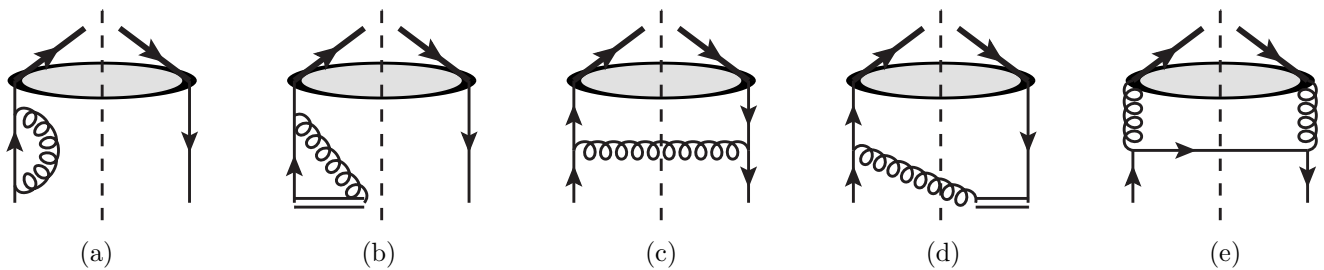}
\\
(a)\hspace{3.3cm}(b)\hspace{3.3cm}(c)\hspace{3.3cm}(d)\hspace{3.3cm}(e)
\end{center}
\caption{\it
One-loop diagrams for the collinear FF and the TMDFF. 
Hermitian conjugates of diagrams (a), (b) and (d) are not shown. 
Double lines stand for collinear Wilson line.
}
\label{fig:collinear}
\end{figure*}

In this appendix we present the calculation of the unpolarized quark-TMDFF at ${\cal O}(\as)$, using dimensional regularization with the $\overline{\rm MS}$-scheme ($\m^2\to\m^2 e^{\g_E}/(4\pi)$) for ultra-violet divergences and the $\D$-regulator~\cite{GarciaEchevarria:2011rb} for IR and rapidity divergences. 
With this regulator, we write the poles of the fermion propagators with a real and positive parameters $\D^\pm$,
\begin{align}\label{eq:fermionsDelta}
\frac{i(\pslash+\kslash)}{(p+k)^2+i0} &\longrightarrow
\frac{i(\pslash+\kslash)}{(p+k)^2+i\D^-}\,,
\nn\\
\quad\quad
\frac{i(\bpslash+\kslash)}{(\bp+k)^2+i0} &\longrightarrow
\frac{i(\bpslash+\kslash)}{(\bp+k)^2+i\D^+}\,,
\end{align}
and for collinear and soft Wilson lines one has
\begin{align}\label{eq:deltas}
\frac{1}{k^{\pm}\pm i0} \longrightarrow
\frac{1}{k^{\pm}\pm i\d^{\pm}}
\,.
\end{align}
Given the fact that the soft and collinear matrix elements should reproduce the soft and collinear limits of full QCD, then they need to be regulated consistently, so $\d^\pm$ are related with $\D^\pm$ through the large components of the collinear fields,
\begin{align} \label{eq:regul_DeltaDY}
\d^+ = \frac{\D^+}{\bp^-}\,, \quad &\quad \quad \d^- = \frac{\D^-}{p^+} 
\,.
\end{align}
Note that $\D^\pm$ (and hence $\d^\pm$) are regulator parameters, and are set to zero unless they regulate any divergence.

Let us now proceed with the calculation.
The unpolarized TMDFF is defined through Eq.~\eqref{eq:TMDFF}:
\begin{align}
D_\bn(z,{\hat P}_{hT};\z_D,\m^2;\D^+) &=
\int d^2\vecb b_\perp\, e^{-i\vecbe b_\perp \cd \vecbe k_{\bn\perp}}\,
\tilde{J}_{\bn}^{(0)}(z,b_T;\m^2;\D^+)\,
\tilde{S}_{+}(b_T;\z_D,\m^2;\D^+)
\,,
\end{align}
where $J_{\bn}^{(0)}=\D^{(0)[\g^-]}=\frac{1}{2}{\rm Tr}(\D^{(0)}\g^-)$ and we average over the spin of the hadron.
This function corresponds to $D_1$ in the spin decomposition in Eq.~\eqref{eq:tmdffdecompositionk}.
Notice that $\tilde{J}_{\bn}^{(0)}$ stands for the pure collinear matrix element, i.e., without any contamination from the soft region. 
However, given the equivalence between the subtraction of the zero-bin and the soft function (with the $\D$-regulator) established in Ref.~\cite{GarciaEchevarria:2011rb}, we can rewrite the previous definition as
\begin{align}
\label{eq:TMDFFwithnaive}
D_\bn(z,{\hat P}_{hT};\z_D,\m^2;\D^+) &=
\int d^2\vecb b_\perp\, e^{-i\vecbe b_\perp \cd \vecbe k_{\bn\perp}}\,
\tilde{J}_{\bn}(z,b_T;Q^2,\m^2;\D^+,\D^-)\,
\tilde{S}_{-}^{-1}(b_T;\z_F,\m^2;\D^-)
\,,
\end{align}
where now $\tilde{J}_{\bn}$ stands for the naively calculated collinear matrix element.
Notice that $\tilde{J}_{\bn}$ depends on $Q^2$, and when combined with the piece of the soft function which depends on $\z_F$, we will recover the correct $\z_D$-dependence. 
Technically this is achieved through the following steps. 
The pure collinear matrix element $\tilde{J}_\bn^{(0)}$ is obtained from the naive collinear one $\tilde{J}_\bn$ through
\begin{align}
\tilde{J}_{\bn}^{(0)}(z,b_T;\m^2;\D^+) &=
\frac{\tilde{J}_{\bn}(z,b_T;Q^2,\m^2;\D^+,\D^-)}{\tilde{S}(b_T;Q^2,\m^2;\D^+,\D^-)}
\,,
\end{align}
and given the splitting of the soft function in Eq.~\eqref{eq:splitting}, then the TMDFF $D_\bn$ can be written as
\begin{align}
D_\bn(z,{\hat P}_{hT};\z_D,\m^2;\D^+) &=
\int d^2\vecb b_\perp\, e^{-i\vecbe b_\perp \cd \vecbe k_{\bn\perp}}\,
\frac{\tilde{J}_{\bn}(z,b_T;Q^2,\m^2;\D^+,\D^-)}
{\tilde{S}_{-}(b_T;\z_F,\m^2;\D^-)\,\tilde{S}_{+}(b_T;\z_D,\m^2;\D^+)}\,
\tilde{S}_{+}(b_T;\z_D,\m^2;\D^+)
\,,
\end{align}
form which we get $D_\bn$ in Eq.~\eqref{eq:TMDFFwithnaive}.
This is the equation that is used below to calculate the TMDFF.

At tree level the collinear matrix element is
\begin{align}
J_{\bn 0} &=
\d(1-z) \d^{(2)}(\vecb k_{\bn\perp})
\,.
\end{align}

The Wave Function Renormalization (WFR) diagram~\ref{fig:collinear}a gives
\begin{align}\label{1a}
i\bpslash \S^{(\ref{fig:collinear}a)}(\bp) &=
-g^2 C_F \d(1-z)\d^{(2)}(\vecb k_{\bn\perp})\m^{2\e} \int \frac{d^dk}{(2\pi)^d}
\frac{-(d-2)(\bpslash - \kslash)}{[(\bp-k)^2+i\D^+][k^2+i0]}
\nn\\
&=
i\bpslash\frac{\alpha_s C_F}{2\pi}\d(1-z)\d^{(2)}(\vecb k_{\bn\perp})
\le[ \frac{1}{2\veuv}+\frac{1}{2}\ln\frac{\m^2}{-i\D^+}+\frac{1}{4} \ri]
\,.
\end{align}
Combined with its Hermitian conjugate  we get
\begin{align}
\S(\bp) &= \S^{(\ref{fig:collinear}a)+(\ref{fig:collinear}a)^*}(\bp) =
\frac{\alpha_s C_F}{2\pi}\d(1-z)\d^{(2)}(\vecb k_{\bn\perp})
\le[ \frac{1}{\veuv}+\ln\frac{\m^2}{\D^+}+\frac{1}{2} \ri]
\,,
\end{align}
which contributes to the TMDFF matrix element with $-\frac{1}{2}\S(\bp)$.

All tadpole diagrams are identically $0$, since $n^2=\nb^2=0$ and they will not be considered any further.
Diagram~\ref{fig:collinear}b and its Hermitian conjugate give
\begin{align}\label{eq:1b}
J_{\bn1}^{(\ref{fig:collinear}b)+(\ref{fig:collinear}b)^*}&=
-2ig^2C_F \d(1-z)\d^{(2)}(\vecb k_{\bn\perp}) \m^{2\e} \int \frac{d^dk}{(2\pi)^d}
\frac{\bp^-+k^-}{[k^-+i\d^-][(\bp+k)^2+i\D^+][k^2+i0]}
+ h.c.
\nn\\
&=
\frac{\a_s C_F}{2\pi}
\d(1-z)\d^{(2)}(\vecb k_{\bn\perp})
\left[
\frac{2}{\veuv}\ln\frac{\d^-}{\bp^-} + \frac{2}{\veuv} - \ln^2\frac{\d^-\D^+}{\bp^-\m^2}
- 2\ln\frac{\D^+}{\m^2} + \ln^2\frac{\D^+}{\m^2} + 2 + \frac{5\pi^2}{12}
\right]
\,.
\end{align}

Diagram~(\ref{fig:collinear}c) gives
\begin{align}\label{eq:1c}
J_{\bn1}^{(\ref{fig:collinear}c)}&=
2\pi g^2 C_F \bp^- \frac{1}{z}\int\frac{d^dk}{(2\pi)^d}
\d(k^2)\theta(-k^-)\frac{2(1-\ve) k_T^2
\d\le((1-1/z)\bp^--k^-\ri) \d^{(2)}(\vecb k_\perp+\vecb k_{\bn\perp})}
{[(\bp-k)^2+i\D^+] [(\bp-k)^2-i\D^+]}
\nn\\
&=
\frac{\a_s C_F}{2\pi^2} \frac{1}{z}\frac{1-z}{z}
\frac{k_{\bn T}^2}{\left| k_{\bn T}^2-i\D^+(1-1/z)\right|^2}
\,,
\end{align}
where we have written $\le|\frac{z-1}{z}\ri|=\frac{1-z}{z}$ with $z \in [0,1]$.

Diagram~(\ref{fig:collinear}d) and its Hermitian conjugate give
\begin{align}\label{eq:1d}
J_{\bn1}^{(\ref{fig:collinear}d)+(\ref{fig:collinear}d)^*}&=
-4\pi g^2 C_F \bp^- \frac{1}{z} \int\frac{d^dk}{(2\pi)^d}
\d(k^2)\theta(-k^-)\frac{(\bp^--k^-)\,
\d\le((1-1/z)\bp^--k^-\ri) \d^{(2)}(\vecb k_\perp+\vecb k_{\bn\perp})}
{[k^--i\d^-][(\bp-k)^2+i\D^+]}
+ h.c.
\nn\\
&=
\frac{\alpha_s C_F}{2\pi^2}
\frac{1}{z^2}
\left[\frac{1}{(1-1/z)-i\d^-/\bp^-}\right]
\left[
\frac{-1}{k_{\bn T}^2-i\D^+(1-1/z)}
\right]
+ h.c.
\,.
\end{align}

Now we list the Fourier transform of the previous results:
\begin{align}
\tilde{J}_{\bn0} &=
\d(1-z)
\,.
\end{align}

\begin{align}
\tilde \S(\bp) &=
\frac{\alpha_s C_F}{2\pi}\d(1-z)
\le[ \frac{1}{\veuv}+\ln\frac{\m^2}{\D^+}+\frac{1}{2} \ri]
\,.
\end{align}

\begin{align}
\tilde {J}_{\bn1}^{(\ref{fig:collinear}b)+(\ref{fig:collinear}b)^*}&=
\frac{\a_s C_F}{2\pi}
\d(1-z)
\left[
\frac{2}{\veuv}\ln\frac{\d^-}{\bp^-} + \frac{2}{\veuv} - \ln^2\frac{\d^-}{\bp^-}
-2\ln\frac{\d^-}{\bp^-}\ln\frac{\D^+}{\m^2}
- 2\ln\frac{\D^+}{\m^2} + 2 + \frac{5\pi^2}{12}
\right]
\,.
\end{align}

\begin{align}
\tilde{J}_{\bn1}^{(\ref{fig:collinear}c)}&=
\frac{\a_s C_F}{2\pi} \frac{1}{z}\frac{1-z}{z}
\ln\frac{4e^{-2\g_E}}{\D^+\frac{1-z}{z}b_T^2}
\,.
\end{align}

\begin{align}
\tilde{J}_{\bn1}^{(\ref{fig:collinear}d)+(\ref{fig:collinear}d)^*}&=
-\frac{\alpha_s C_F}{2\pi}
\frac{1}{z^2}
\left[\frac{1}{(1-1/z)-i\d^-/\bp^-}\right]
\ln\frac{4e^{-2\g_E}}{i\D^+\frac{1-z}{z}b_T^2}
+ h.c.
\nn\\
&=
\frac{\as C_F}{2\pi} \le[
\ln\frac{4e^{-2\g_E}}{\D^+b_T^2}
\le(
\frac{2/z}{(1-z)_+} - 2\ln\frac{\d^-}{\bp^-}\d(1-z)
\ri)
-\frac{2}{z}\le(\frac{\ln(1-z)}{1-z}\ri)_+ 
+ \le(\ln^2\frac{\d^-}{\bp^-}+\frac{\pi^2}{12}\ri)\d(1-z)
\ri.
\nn\\
&\le.
+ \frac{(2/z)\,\ln z}{(1-z)_+} 
- \frac{\pi^2}{2}\d(1-z)
\ri]
\,.
\end{align}

We have used the following identity in $d=2-2\ve$ to perform the Fourier transforms:
\begin{align}
\int d^d\vecb k_\perp e^{i\vecbe k_\perp\cdot \vecbe b_\perp}
f(k_T)
&=
b_T^{-d} (2\pi)^\frac{d}{2} \int_0^\infty dy\, y^\frac{d}{2} J_{\frac{d}{2}-1}(y)\,
f\left(\frac{y}{b_T}\right)\,,
\end{align}
with the particular results
\begin{align}
\int d^d\vecb k_\perp e^{i\vecbe k_\perp\cdot \vecbe b_\perp}
\frac{1}{k_T^2-i\L^2}
&=
\pi\, \ln\frac{4e^{-2\g_E}}{-i\L^2 b_T^2}\,,
\nn\\
\int d^d\vecb k_\perp e^{i\vecbe k_\perp\cdot \vecbe b_\perp}
\frac{k_T^2}{k_T^4+\L^4}
&=
\pi\, \ln\frac{4e^{-2\g_E}}{\L^2 b_T^2}
\,,
\end{align}
when $\L\to 0$.

We have also used the following relations:
\begin{align}
f(z)\left[\frac{1}{(1-1/z)-i\d^-/\bp^-}
+\frac{1}{(1-1/z)+i\d^-/\bp^-}\right] &=
f(z)\le[\frac{-2z}{(1-z)_+} + 2\ln\frac{\d^-}{\bp^-}\d(1-z)\ri]
\,,
\nn\\
f(z)\left[\frac{\ln(1-z)}{(1-1/z)-i\d^-/\bp^-}
+\frac{\ln(1-z)}{(1-1/z)+i\d^-/\bp^-}\right] &=
f(z)\le[-2z\le(\frac{\ln(1-z)}{1-z}\ri)_+ +
\le(\ln^2\frac{\d^-}{\bp^-}+\frac{\pi^2}{12}\ri)\d(1-z)
\ri]
\,,
\nn\\
f(z)\left[\frac{1}{(1-1/z)-i\d^-/\bp^-}-\frac{1}{(1-1/z)+i\d^-/\bp^-}\right] &=
i\pi\d(1-z) f(z)
\,,
\end{align}
where $f(z)$ is any function regular at $z\to 1$.
Similar relations were also used to derive some of the results below, however they will not be displayed.

Thus, in IPS, the collinear matrix element, for the partonic channel of a quark fragmenting into a quark is
\begin{align}
\tilde{J}_{\bn,q\leftarrow q}(z,b_T) &=
\d(1-z) + \frac{\as C_F}{2\pi} \le\{
\d(1-z)\le[
\frac{2}{\veuv}\ln\frac{\D^-}{Q^2} + \frac{3}{2\veuv}\ri]
\ri.
\nn\\
&\le.
-L_\perp \le(\frac{1}{z^2}{\cal P}_{q\leftarrow q}-\frac{3}{2}\d(1-z)\ri)
+2L_\perp\ln\frac{\D^-}{Q^2}\d(1-z) - \frac{1-z}{z^2}\ln\frac{1-z}{z}
-\frac{2}{z}\le(\frac{\ln(1-z)}{1-z}\ri)_+
+ \frac{(2/z)\,\ln z}{(1-z)_+}
\ri.
\nn\\
&\le.
-\ln\frac{\D^+}{\m^2}\le( \frac{1}{z^2}{\cal P}_{q\leftarrow q} \ri)
+\frac{7}{4}\d(1-z)
\ri\}
\,,
\end{align}
where $L_\perp=\ln(\m^2 b_T^2 e^{2\g_E}/4)$ and, to ${\cal{O}}(\alpha_s)$, the DGLAP kernel ${\cal P}_{q\leftarrow q}$ is the same as for the collinear PDF:
\begin{align}
{\cal P}_{q\leftarrow q} &= 
\left( \frac{1+z^2}{1-z} \right)_+ =
\frac{1+z^2}{(1-z)_+} + \frac{3}{2}\d(1-z) =
\frac{2z}{(1-z)_+} + (1-z) + \frac{3}{2}\d(1-z)
\,.
\end{align}
As in the case of TMDPDF~\cite{GarciaEchevarria:2011rb} the mixed divergences are a signal of rapidity divergences which needs to be eliminated to get a well-defined hadronic quantities.

\begin{figure*}[t]
\begin{center}
\includegraphics[width=0.4\textwidth]{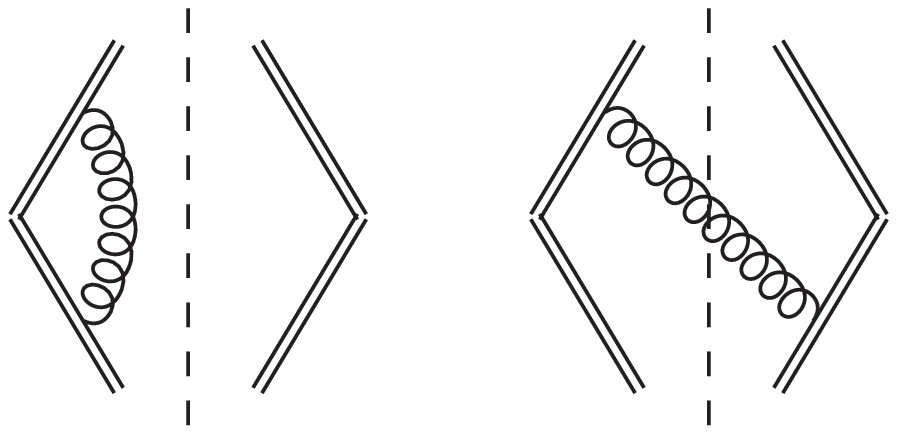}
\\
(a)\hspace{3.6cm}(b)
\end{center}
\caption{\it
One-loop diagrams for the soft function. 
Hermitian conjugate of diagrams (a) and (b) are not shown. 
Double lines stand for soft Wilson lines.
}
\label{fig:soft}
\end{figure*}

Let us now calculate the soft function.
Diagram~(\ref{fig:soft}a) and its Hermitian conjugate give
\begin{align}\label{eq:soft_virtual}
S_1^{(\ref{fig:soft}a)+(\ref{fig:soft}a)^*}&=
-2ig^2 C_F \d^{(2)}(\vecb k_{s\perp}) \mu^{2 \eps}
\int \frac{d^d k}{(2 \pi)^d} \frac{1}{[k^-+i\d^-] [k^++i\d^+] [k^2+i0]} +h.c.
\nn\\
&=
-\frac{\alpha_s C_F}{2\pi}
\d^{(2)}(\vecb k_{s\perp})
\left[\frac{2}{\veuv^2} - \frac{2}{\veuv}\ln\frac{\d^+\d^-}{\mu^2} 
+ \ln^2\frac{\d^+\d^-}{\mu^2} - \frac{\pi^2}{2}\right]\, .
\end{align}

And for diagram~(\ref{fig:soft}b) and its Hermitian conjugate we have
\begin{align}\label{eq:soft_real}
S_1^{(\ref{fig:soft}b)+(\ref{fig:soft}b)^*}&=
-4\pi g^2 C_F  \int\frac{d^dk}{(2\pi)^d}
\d^{(2)}(\vecb k_\perp+\vecb k_{s\perp}) \d(k^2)\theta(-k^-) \frac{1}{[k^--i\d^-][-k^++i\d^+]} + h.c.
\nn\\
&=
-\frac{\a_s C_F}{\pi^2}
\frac{1}{k_{sT}^2+\d^+\d^-} \ln\frac{\d^+\d^-}{k_{sT}^2}
\,.
\end{align}

In IPS we then get
\begin{align}
\tilde S_1^{(\ref{fig:soft}a)+(\ref{fig:soft}a)^*}&=
-\frac{\alpha_s C_F}{2\pi}
\left[\frac{2}{\veuv^2} - \frac{2}{\veuv}\ln\frac{\d^+\d^-}{\mu^2} 
+ \ln^2\frac{\d^+\d^-}{\mu^2} - \frac{\pi^2}{2}\right]
\,,
\end{align}
and
\begin{align}
\tilde S_1^{(\ref{fig:soft}b)+(\ref{fig:soft}b)^*}&=
\frac{\a_s C_F}{2\pi}
\le( \ln^2\frac{4e^{-2\g_E}}{\d^+\d^- b_T^2} - \frac{\pi^2}{3} \ri)
\,,
\end{align}
where we  have used the following:
\begin{align}
\int d^d\vecb k_\perp e^{i\vecbe k_\perp\cdot \vecbe b_\perp}
\frac{1}{k_T^2+\L^2} \ln\frac{\L^2}{k_T^2}
&=
\pi \left(
-\frac{1}{2}\ln^2\frac{4e^{-2\g_E}}{\L^2 b_T^2} + \frac{\pi^2}{6}
\right)
\,.
\end{align}

The complete soft function in IPS at ${\cal O}(\as)$ is
\begin{align}
\tilde S &=
1 + \frac{\as C_F}{2\pi} \le[
-\frac{2}{\veuv^2} + \frac{2}{\veuv}\ln\frac{\D^+\D^-}{\mu^2Q^2}
+L_\perp^2 + 2L_\perp\ln\frac{\D^+\D^-}{\m^2Q^2} + \frac{\pi^2}{6}
\ri]
\,,
\end{align}
which is the same as for DY kinematics (see Eq.~(18) in Ref.~\cite{Echevarria:2012js}).

Combining the collinear and soft matrix elements as in Eq.~\eqref{eq:TMDFFwithnaive} we get the TMDFF in IPS for the $(q\leftarrow q)$ channel:
\begin{align}\label{eq:tmdfffirstorderb}
&\tilde D_{\bn,q\leftarrow q}(z,b_T;\z_\bn,\m^2) = 
\d(1-z) + \le[\tilde {J}_{\bn1,q\leftarrow q} - \frac{1}{2}\d(1-z)\tilde S_1\le(\frac{\D^-}{p^+},\a\frac{\D^-}{\bp^-}\ri)\ri] =
\nn\\
&=
\d(1-z) + \frac{\as C_F}{2\pi} \le\{
\d(1-z)\le[
\frac{1}{\veuv^2}-\frac{1}{\veuv}\ln\frac{\z_\bn}{\m^2}+\frac{3}{2\veuv}\ri]
\ri.
\nn\\
&\le.
-\frac{1}{2}L_\perp^2\d(1-z)
-L_\perp \le(\frac{1}{z^2}{\cal P}_{q\leftarrow q}-\frac{3}{2}\d(1-z)\ri)
-L_\perp\ln\frac{\z_\bn}{\m^2}\d(1-z) + \frac{1-z}{z^2}
+ \frac{\ln z^2}{z^2}{\cal P}_{q\leftarrow q}
-\frac{\pi^2}{12}\d(1-z)
\ri.
\nn\\
&\le.
-\frac{1}{z^2}{\cal P}_{q\leftarrow q}\ln\frac{\D^+}{\m^2}
+\frac{7}{4}\d(1-z)
- \frac{(1-z)}{z^2}\left[1+\ln\frac{1-z}{z}\right] 
-\frac{2}{z}\le(\frac{\ln(1-z)}{1-z}\ri)_+
+ \frac{(2/z)\,\ln z}{(1-z)_+}
- \frac{\ln z^2}{z^2}{\cal P}_{q\leftarrow q}
\ri\}
\,,
\end{align}
where all rapidity divergences are cancelled, as expected.

There is another channel yet to be calculated, i.e., the one that corresponds to a quark fragmenting into a gluon.
If we choose to sum over the physical (transverse) polarizations for the gluons, then we need to calculate one sole diagram, given in Fig.~(\ref{fig:collinear}e).
Thus, we have
\begin{align}
J_{\bn1,g\leftarrow q}^{(\ref{fig:collinear}e)}&=
-g^2 C_F \frac{1}{z} \int\frac{d^4k}{(2\pi)^4}
\d(k^--(1-1/z)\bp^-) \d^{(2)}(\vecb k_\perp+\vecb k_{\bn\perp})
\d(k^2)\theta(-k^-)
\frac{1}{2}\sum_{pol} \e^{\perp*}_\m \e^\perp_\n
\frac{{\rm tr}\le[(\bpslash-\kslash)\g^\m\kslash\g^\n(\bpslash-\kslash)\frac{\nslash}{2}\ri]}{\le[(\bp-k)^2+i\D^+\ri] \le[(\bp-k)^2-i\D^+\ri]}
\nn\\
&=
\frac{\alpha_s C_F}{2\pi^2}
\frac{1}{z^3}
\frac{\le[(1-\ve)z^2+2(1-z)\ri]k_{\bn T}^2}
{\le[k_{\bn T}^2-i\D^+(1-1/z)\ri]\le[k_{\bn T}^2+i\D^+(1-1/z)\ri]}
\,,
\end{align}
where we have used the following:
\begin{align}
\sum_{pol} \e^{\perp*}_\m \e^\perp_\n &=
- g_{\m\n}^\perp
=
g_{\m\n}-\frac{1}{2}(n^\m\bn^\n + n^\n\bn^\m)
\,.
\end{align}
The Fourier transform on this result is
\begin{align}
\tilde{J}_{\bn1,g\leftarrow q}^{(\ref{fig:collinear}e)}&=
\frac{\alpha_s C_F}{2\pi}
\frac{1}{z^2}{\cal P}_{g\leftarrow q}
\ln\frac{4e^{-2\g_E}}{\D^+b_T^2(1-z)/z}
\,,
\end{align}
where the DGLAP splitting function of a quark into a gluon is
\begin{align}
{\cal P}_{g\leftarrow q} &=
\frac{z^2+2(1-z)}{z}
\,.
\end{align}
Given the fact that there is no soft contribution for this channel, the TMDFF in IPS and for the $(g\leftarrow q)$ partonic channel is then,
\begin{align}\label{eq:tmdfffirstorderb_qg}
&\tilde D_{\bn,g\leftarrow q}(z,b_T;\z_\bn,\m^2) = 
\frac{\alpha_s C_F}{2\pi}
\frac{1}{z^2}{\cal P}_{g\leftarrow q}
\ln\frac{4e^{-2\g_E}}{\D^+b_T^2(1-z)/z}
\,.
\end{align}

\section{Refactorization of TMDFF in terms of collinear FF}
\label{sec:tmdffope}

Below we perform an OPE of the quark-TMDFF onto the integrated quark FF and calculate the matching coefficient at ${\cal O}(\as)$.
The OPE of the renormalized TMDFF onto the renormalized  FF is
\begin{align}\label{eq:opetmdff}
\tilde D_\bn^R(z,b_T;\z_\bn,\m^2) &=
\int_z^1 \frac{d\hat{z}}{\hat{z}^{3-2\ve}} 
\tilde C_\bn\le(\frac{z}{\hat{z}},b_T;\z_\bn,\m^2\ri)
d_\bn^R(\hat{z};\m)
\,,
\end{align}
where the FF is defined as~\cite{Collins:1981uw}
\begin{align}\label{eq:ffdef}
d_\bn(z;\m)  &=
\frac{z^{d-3}}{2} \int\frac{dy^+}{2\pi}
e^{i\frac{1}{2}y^+\bp^-/z}
\frac{1}{2} \sum_s \sum_{X}\, tr\,
\frac{\nslash}{2}
\sandwich{0}{\le[\tilde W_\bn^{\dagger} \x_\bn\ri](y^+,0^-,\vec 0_\perp)}{\bP s,X}
\sandwich{\bP s,X}{\le[\bar\x_\bn \tilde W_\bn\ri](0)}{0}
\,.
\end{align}
One can easily verify that the matching coefficient at one loop is
\begin{align}\label{eq:opecoeff}
\tilde C_{\bn1} &=
\tilde D_{\bn1}^R - \frac{d_{\bn1}^R}{z^{2-2\ve}}
\,,
\end{align}

where the TMDFF is given in Eq.~(\ref{eq:tmdfffirstorderb}).
In order to properly obtain the matching coefficient, we need to calculate the collinear FF at ${\cal O}(\as)$ (while using  the same regulators used for the TMDFF).

At tree level, the FF  is
\begin{align}
d_{\bn 0} &=
\frac{z^{1-2\ve}}{2} \int\frac{dy^+}{2\pi}
e^{i\frac{1}{2}y^+\bp^-/z}
\frac{1}{2} \sum_s {\rm tr} \frac{\nslash}{2}
\sandwich{0}{\x_\bn(y^+,0^-,\vec 0_{\perp})}{\bp}
\sandwich{\bp}{\bar \x_\bn(0)}{0}
\nn\\
&=
\d(1-z)
\,.
\end{align}
 
As before, the WFR contribution is $-\frac{1}{2}\S(\bp)$.

Diagram~(\ref{fig:collinear}b) and its Hermitian conjugate give
\begin{align}
\label{eq:ffn_virtuals1b}
d_{\bn1}^{(\ref{fig:collinear}b)+(\ref{fig:collinear}b)^*}&=
-2ig^2C_F \d(1-z)\m^{2\e} 
\int \frac{d^dk}{(2\pi)^d}
\frac{\bp^-+k^-}{[k^-+i\d^-][(\bp+k)^2+i\D^+][k^2+i0]}
+ h.c.
\nn\\
&=
\frac{\a_s C_F}{2\pi}\d(1-z)
\left[
\frac{2}{\veuv}\ln\frac{\d^-}{\bp^-} + \frac{2}{\veuv} 
- \ln^2\frac{\d^-}{\bp^-} - 2\ln\frac{\d^-}{\bp^-}\ln\frac{\D^+}{\m^2}
- 2\ln\frac{\D^+}{\m^2} + 2 + \frac{5\pi^2}{12}
\right]
\,.
\end{align}

The contribution of diagram~(\ref{fig:collinear}c) is
\begin{align}
d_{\bn1}^{(\ref{fig:collinear}c)}&=
2\pi g^2 C_F \bp^- z^{1-2\ve} \m^{2\ve}\int\frac{d^dk}{(2\pi)^d}
\d(k^2)\theta(-k^-)
\frac{-2(1-\ve) k_\perp^2\d\le((1-1/z)\bp^--k^-\ri)}{[(\bp-k)^2+i\D^+][(\bp-k)^2-i\D^+]}
\nn\\
&=
\frac{\as C_F}{2\pi} (1-z) \left[ \frac{1}{\veuv} + \ln\frac{\m^2}{\D^+} 
- 1 - \ln\frac{1-z}{z} - \ln z^2\right]
\,.
\end{align}

Diagram~(\ref{fig:collinear}d) and its Hermitian conjugate give
\begin{align}
d_{\bn1}^{(\ref{fig:collinear}d)+(\ref{fig:collinear}d)^*}&=
-4\pi g^2 C_F \bp^- z^{1-2\ve} \m^{2\ve}
\int\frac{d^dk}{(2\pi)^d}
\d(k^2)\theta(-k^-)
\frac{(\bp^--k^-)\d\le((1-1/z)\bp^--k^-\ri)}{[k^--i\d^-][(\bp-k)^2+i\D^+]}
+ h.c.
\nn\\
&=
\frac{\as C_F}{2\pi} \left[
\left( \frac{1}{\veuv} + \ln\frac{\m^2}{\D^+} \right)
\left( \frac{2z}{(1-z)_+} - 2\d(1-z)\ln\frac{\d^-}{\bp^-} \right)
-\frac{2z\ln z^2}{(1-z)_+}
-2z\le(\frac{\ln(1-z)}{1-z}\ri)_+ 
\ri.
\nn\\
&
\le.
+\le(\ln^2\frac{\d^-}{\bp^-}+\frac{\pi^2}{12}\ri)\d(1-z)
+\frac{2z\ln z}{(1-z)_+}
- \frac{\pi^2}{2} \d(1-z)
\right]\,,
\end{align}
where we have used the following relations:
\begin{align}\label{eq:distributions}
&\frac{(-1+1/z)^{-\ve}z^{-2\ve}}{(1-1/z)-i\d^-/\bp^-} 
+ \frac{(-1+1/z)^{-\ve}z^{-2\ve}}{(1-1/z) + i\d^-/\bp^-}  
=
-\frac{2z}{(1-z)_+} + 2\d(1-z)\ln\frac{\d^-}{\bp^-}
\nn\\
&
- \ve\le[
-\frac{2z\ln z^2}{(1-z)_+}
-2z\le(\frac{\ln(1-z)}{1-z}\ri)_+ 
+\le(\ln^2\frac{\d^-}{\bp^-}+\frac{\pi^2}{12}\ri)\d(1-z)
+\frac{2z\ln z}{(1-z)_+}
\ri]
+{\cal O}(\varepsilon^2)\,,
\nn\\
&
\frac{(-1+1/z)^{-\ve}z^{1-2\ve}}{(z-1)-iz\d^-/\bp^-}
- \frac{(-1+1/z)^{-\ve}z^{1-2\ve}}{(z-1)+iz\d^-/\bp^-} =
i\pi\d(1-z)
+{\cal O}(\varepsilon)
\,.
\end{align}

Combining the virtual and real contributions we get the collinear FF to first order in $\as$,
\begin{align}
\label{eq:ffdelta}
d_{\bn}(z;\m) &= 
\delta(1-z)
+\frac{\as C_F}{2\pi} \left[
{\cal P}_{q\leftarrow q} \left( \frac{1}{\veuv} - \ln\frac{\D^+}{\m^2}\right)
\ri.
\nn\\
&\le.
+ \frac{7}{4}\d(1-z) - (1-z)\left[1+\ln\frac{1-z}{z}\right]
-2z\le(\frac{\ln(1-z)}{1-z}\ri)_+ + \frac{2z\ln z}{(1-z)_+} 
- \ln z^2 {\cal P}_{q\leftarrow q}
\right]
\,,
\end{align}
where we have used that
\begin{align}
\ln z^2{\cal P}_{q\leftarrow q} &= 
\frac{2z\ln z^2}{(1-z)_+} + (1-z)\ln z^2
\,.
\end{align}

Taking the renormalized one-loop result for the FF, we get
\begin{align}
\frac{d_{\bn1}^R}{z^{2-2\ve}} &= 
\frac{\as C_F}{2\pi} \le[
-\frac{1}{z^2}{\cal P}_{q\leftarrow q}\ln\frac{\D^+}{\m^2}
+ \frac{7}{4}\d(1-x) - \frac{(1-z)}{z^2}\left[1+\ln\frac{1-z}{z}\right]
-\frac{2}{z}\le(\frac{\ln(1-z)}{1-z}\ri)_+
+\frac{(2/z)\ln z}{(1-z)_+} - \frac{\ln z^2}{z^2}{\cal P}_{q\leftarrow q}
\ri]
\,,
\end{align}
where the factor $z^{2\ve}$ has no effect given that the renormalized FF, $d_{\bn}^R$, has no poles.
Combining this result with the TMDFF at one loop~\footnote{Remember that for the OPE we need to consider the renormalized TMDFF.}, given in Eq.~\eqref{eq:tmdfffirstorderb}, as it appears in Eq.~\eqref{eq:opecoeff}, we get the matching coefficient at ${\cal O}(\as)$,
\begin{align}\label{eq:coeffquark}
\tilde{C}_{\bn}(z,b_T;\z_\bn,\m^2) &=
\d(1-z) 
+ 
\frac{\as C_F}{2\pi}\le[
-\frac{1}{2}L_\perp^2\d(1-z)
-L_\perp \le(\frac{1}{z^2}{\cal P}_{q\leftarrow q} - \frac{3}{2}\d(1-z)\ri)
\ri.
\nn\\
&\le.
-L_\perp\ln\frac{\z_\bn}{\m^2}\d(1-z) 
+ \frac{1-z}{z^2} + \frac{\ln z^2}{z^2}{\cal P}_{q\leftarrow q}
-\frac{\pi^2}{12}\d(1-z)
\ri]
\,.
\end{align}
This result is consistent with the one calculated in Ref.~\cite{Collins:2011zzd}, apart from the $\pi^2$-term, which is related to a different convention for the $\overline{\rm MS}$-scheme (see Sec.~VI in Ref.~\cite{Collins:2012uy}).
Notice that this coefficient, as the one for the OPE of the TMDPDF onto the collinear PDF derived in Ref.~\cite{GarciaEchevarria:2011rb} depends explicitly on $Q^2$ (through $\z_\bn$).
This dependence can be exponentiated by following the same steps as in the case of the TMDPDF, thus we are  able to write the TMDFF as
\begin{align}\label{eq:tmdffq2}
\tilde{D}_\bn(z,b_T;\z_\bn,\m^2) &= 
\le( \frac{\z_\bn b_T^2}{4e^{-2\g_E}} \ri)^{-D(b_T;\m)} 
\tilde{C}_\bn^{\Qslash}(z,b_T;\m^2)
\otimes d_\bn(z;\m)
\,,
\end{align}
where
\begin{align}
\tilde{C}_\bn^{\Qslash}(z,b_T;\m^2) &=
\delta(1-z)+\frac{\as C_F}{2 \pi}\left[
-\frac{1}{z^2}{\cal P}_{q\leftarrow q} L_\perp + \frac{1-z}{z^2} 
- \delta(1-z)
\left(
- \frac{1}{2}L_\perp^2 + \frac{3}{2}L_\perp + \frac{\pi^2}{12}\right)\right]
\,.
\end{align}
and the $D$ function is related to the cusp anomalous dimension through
\begin{align}
\frac{d D}{d\ln\m} &=
\G_{\rm cusp}
\,.
\end{align}

Finally, the contribution to the collinear FF coming from the quark-gluon channel is
\begin{align}
d_{\bn1,g\leftarrow q}^{(\ref{fig:collinear}e)}&=
-g^2 C_F z^{1-2\ve} \m^{2\ve}\int\frac{d^dk}{(2\pi)^d}
\d(k^--(1-1/z)\bp^-) \d(k^2)\theta(-k^-)
\frac{1}{2}\sum_{pol} \e^{\perp*}_\m \e^\perp_\n
\frac{{\rm tr}\le[(\bpslash-\kslash)\g^\m\kslash\g^\n(\bpslash-\kslash)\frac{\nslash}{2}\ri]}{\le[(\bp-k)^2+i\D^+\ri] \le[(\bp-k)^2-i\D^+\ri]}
\nn\\
&=
\frac{\alpha_s C_F}{2\pi}
\le[
{\cal P}_{g\leftarrow q}\le(\frac{1}{\veuv}+\ln\frac{\m^2}{\D^+}\ri)
- \ln[z(1-z)] {\cal P}_{g\leftarrow q} - z
\ri]
\,.
\end{align}
Combining this result with the one of the TMDFF as in Eq.~\eqref{eq:opecoeff} we get for the coefficient in this channel:
\begin{align}\label{eq:coeffgluon}
\tilde{C}_{\bn,g\leftarrow q} &=
\frac{\as C_F}{2\pi}
\le[
-L_\perp\frac{1}{z^2}{\cal P}_{g\leftarrow q}
+ \frac{\ln z^2}{z^2}{\cal P}_{g\leftarrow q} + \frac{1}{z}
\ri]
\,.
\end{align}

\section{Hard Part at ${\cal O}(\as)$}
\label{app:hard}

Here we obtain  the hard coefficient contributing to the factorized hadronic tensor of SIDIS. This contribution has to be the same as the for inclusive DIS since the hard part is obtained from virtual Feynman diagrams and, thus, is independent of $q_T$ or $b_T$ and it is only a polynomial in $\log(Q^2/\mu^2)$ and no other parameter-dependence is allowed.  
The hard matching coefficient is calculated by subtracting the effective theory contributions from the hadronic tensor calculated in full QCD.
In IPS, The hadronic tensor at ${\cal O}(\as)$  can be written in terms of the naive collinear (i.e., before any soft or zero-bin subtraction is performed. In the next equation this is taken care off by the ``$-$'' sign in front of the soft function) and soft matrix element as
\begin{align}\label{eq:hadronictensordis}
\tilde{W}(x,z,b_T;Q) &=
H(Q^2/\mu^2)
\left[
\d(1-x)\d(1-z)
\right.
\nn\\
&\left.
+
\Big(
\d(1-z)\, \tilde{J}_{n1}(x,b_T;Q,\m) +
\d(1-x)\, \tilde{J}_{\bn 1}(z,b_T;Q,\m)
- \d(1-x)\d(1-z)\tilde{S}_1(b_T;Q)
\Big)
\right]
+O(\a_s^2)
\,.
\end{align}
In QCD, the virtual part of $\tilde{W}$ (with the $\D$-regulator) can be computed by calculating the vertex diagram (and the WFR) for DIS kinematics.
Setting $\D^\pm=\D$ for simplicity, the result is
\begin{align}\label{eq:mqcddisv}
\tilde{W}_{QCD}^v &=
\d(1-x)\d(1-z)\le\{
1+
\frac{\a_s C_F }{2\pi}
\le[ - 2\ln^2\frac{\D}{Q^2} 
- 3\ln\frac{\D}{Q^2} 
- \frac{9}{2} + \frac{\pi^2}{2} \ri]
\ri\}
\,.
\end{align}

Collecting the results from Appendix~\ref{app:tmdff_nlo}, we can write the virtual part of the naive collinear and soft matrix elements,
\begin{align}\label{eq:tmdprelimv}
J_{n1}^v &=
\frac{\a_s C_F}{2\pi} \d(1-x)
\left[
\frac{2}{\veuv}\ln\frac{\D}{Q^2} + \frac{3}{2\veuv} 
- \ln^2\frac{\D^2}{Q^2\m^2}
- \frac{3}{2}\ln\frac{\D}{\m^2} + \ln^2\frac{\D}{\m^2} + \frac{7}{4} 
+ \frac{5\pi^2}{12}
\right]
\,,
\nn\\
J_{\bn 1}^v &=
\frac{\a_s C_F}{2\pi} \d(1-z)
\left[
\frac{2}{\veuv}\ln\frac{\D}{Q^2} + \frac{3}{2\veuv} 
- \ln^2\frac{\D^2}{Q^2\m^2}
- \frac{3}{2}\ln\frac{\D}{\m^2} + \ln^2\frac{\D}{\m^2} + \frac{7}{4} 
+ \frac{5\pi^2}{12}
\right]
\,,
\nn\\
S_1^v &=
\frac{\alpha_s C_F}{2\pi}
\left[-\frac{2}{\veuv^2}+\frac{2}{\veuv}\ln\frac{\D^2}{Q^2\mu^2}
-\ln^2\frac{\D^2}{Q^2\mu^2}+\frac{\pi^2}{2}\right]
\,.
\end{align}

Thus, inserting the results above in Eq.~(\ref{eq:hadronictensordis}), the total virtual part of the hadronic tensor $W$ in the effective theory is
\begin{align}\label{eq:mscetdisv}
\tilde{W}_{SCET}^v &=
H(Q^2/\mu^2)
\d(1-x)\d(1-z) \le\{
1+
\frac{\a_s C_F}{2\pi}
\left[
\frac{2}{\veuv^2} 
+ \frac{1}{\veuv} \left( 3 + 2\ln\frac{\m^2}{Q^2}  \right) 
\ri.\ri.
\nn\\
&\le.\le.
- 2\ln^2\frac{\D}{Q^2} - 3\ln\frac{\D}{Q^2}
+ 3\ln\frac{\m^2}{Q^2} 
+ \ln^2\frac{\m^2}{Q^2} + \frac{7}{2} 
+ \frac{\pi^2}{3}
\ri]\ri\}
\,,
\end{align}
where the UV divergences are canceled by a standard renormalization process. 
Notice that the IR (the $\Delta$-dependent terms) contributions in Eqs.~(\ref{eq:mqcddisv}) and~(\ref{eq:mscetdisv}) are the same 9otherwise the matching procedure breaks down).
Thus the matching coefficient between QCD and the effective theory at scale $Q$ is
\begin{align}
H(Q^2/\mu^2) =
1 + \frac{\a_s C_F}{2\pi} \left[
- 3\ln\frac{\m^2}{Q^2} - \ln^2\frac{\m^2}{Q^2} - 8 + \frac{\pi^2}{6}
\right]\,.
\end{align}
The above result was first derived in Ref.~\cite{Manohar:2003vb}.
At two loops one can find it in Ref.~\cite{Idilbi:2006dg}.

It is worth emphasizing that the hard part, as a Wilson coefficient, cannot depend on any regulator.
In other words, if both theories above and below the matching scale have the same IR physics, then the matching coefficient has to be free from any IR regulator.
However, even if this is the case, since the hard part is calculated by a subtraction method, i.e., subtracting the contribution of the theory below the matching scale from the one of the theory above it,  it is crucial to regulate the divergences in both theories consistently, i.e., one needs to employ the same regulator in the effective theory (or in the building blocks of the factorized hadronic tensor) as well as in full QCD.
Otherwise the cancellation of divergences in perturbation theory is not realized properly.
One can find several examples in the literature where this is actually the case, see e.g. Refs.~\cite{Ji:2004wu,Chen:2006vd,Kang:2011mr,Ma:2012hh,Sun:2013hua}, ending up with a hard part which actually do depend on a non-ultraviolet  regulator. 
\bibliography{references}

\begin{thebibliography}{44}%
\makeatletter
\providecommand \@ifxundefined [1]{%
 \@ifx{#1\undefined}
}%
\providecommand \@ifnum [1]{%
 \ifnum #1\expandafter \@firstoftwo
 \else \expandafter \@secondoftwo
 \fi
}%
\providecommand \@ifx [1]{%
 \ifx #1\expandafter \@firstoftwo
 \else \expandafter \@secondoftwo
 \fi
}%
\providecommand \natexlab [1]{#1}%
\providecommand \enquote  [1]{``#1''}%
\providecommand \bibnamefont  [1]{#1}%
\providecommand \bibfnamefont [1]{#1}%
\providecommand \citenamefont [1]{#1}%
\providecommand \href@noop [0]{\@secondoftwo}%
\providecommand \href [0]{\begingroup \@sanitize@url \@href}%
\providecommand \@href[1]{\@@startlink{#1}\@@href}%
\providecommand \@@href[1]{\endgroup#1\@@endlink}%
\providecommand \@sanitize@url [0]{\catcode `\\12\catcode `\$12\catcode
  `\&12\catcode `\#12\catcode `\^12\catcode `\_12\catcode `\%12\relax}%
\providecommand \@@startlink[1]{}%
\providecommand \@@endlink[0]{}%
\providecommand \url  [0]{\begingroup\@sanitize@url \@url }%
\providecommand \@url [1]{\endgroup\@href {#1}{\urlprefix }}%
\providecommand \urlprefix  [0]{URL }%
\providecommand \Eprint [0]{\href }%
\providecommand \doibase [0]{http://dx.doi.org/}%
\providecommand \selectlanguage [0]{\@gobble}%
\providecommand \bibinfo  [0]{\@secondoftwo}%
\providecommand \bibfield  [0]{\@secondoftwo}%
\providecommand \translation [1]{[#1]}%
\providecommand \BibitemOpen [0]{}%
\providecommand \bibitemStop [0]{}%
\providecommand \bibitemNoStop [0]{.\EOS\space}%
\providecommand \EOS [0]{\spacefactor3000\relax}%
\providecommand \BibitemShut  [1]{\csname bibitem#1\endcsname}%
\let\auto@bib@innerbib\@empty
\bibitem [{\citenamefont {Boer}\ and\ \citenamefont
  {Mulders}(1998)}]{Boer:1997nt}%
  \BibitemOpen
  \bibfield  {author} {\bibinfo {author} {\bibfnamefont {D.}~\bibnamefont
  {Boer}}\ and\ \bibinfo {author} {\bibfnamefont {P.}~\bibnamefont {Mulders}},\
  }\href {\doibase 10.1103/PhysRevD.57.5780} {\bibfield  {journal} {\bibinfo
  {journal} {Phys.Rev.}\ }\textbf {\bibinfo {volume} {D57}},\ \bibinfo {pages}
  {5780} (\bibinfo {year} {1998})},\ \Eprint
  {http://arxiv.org/abs/hep-ph/9711485} {arXiv:hep-ph/9711485 [hep-ph]}
  \BibitemShut {NoStop}%
\bibitem [{\citenamefont {Collins}(2011)}]{Collins:2011zzd}%
  \BibitemOpen
  \bibfield  {author} {\bibinfo {author} {\bibfnamefont {J.}~\bibnamefont
  {Collins}},\ }\href@noop {} {\emph {\bibinfo {title} {{Foundations of
  perturbative QCD}}}}\ (\bibinfo  {publisher} {Cambridge monographs on
  particle physics, nuclear physics and cosmology. 32},\ \bibinfo {year}
  {2011})\BibitemShut {NoStop}%
\bibitem [{\citenamefont {Echevarria}\ \emph
  {et~al.}(2013{\natexlab{a}})\citenamefont {Echevarria}, \citenamefont
  {Idilbi},\ and\ \citenamefont {Scimemi}}]{Echevarria:2012js}%
  \BibitemOpen
  \bibfield  {author} {\bibinfo {author} {\bibfnamefont {M.~G.}\ \bibnamefont
  {Echevarria}}, \bibinfo {author} {\bibfnamefont {A.}~\bibnamefont {Idilbi}},
  \ and\ \bibinfo {author} {\bibfnamefont {I.}~\bibnamefont {Scimemi}},\ }\href
  {\doibase 10.1016/j.physletb.2013.09.003} {\bibfield  {journal} {\bibinfo
  {journal} {Phys.Lett.}\ }\textbf {\bibinfo {volume} {B726}},\ \bibinfo
  {pages} {795} (\bibinfo {year} {2013}{\natexlab{a}})},\ \Eprint
  {http://arxiv.org/abs/1211.1947} {arXiv:1211.1947 [hep-ph]} \BibitemShut
  {NoStop}%
\bibitem [{\citenamefont {Aybat}\ and\ \citenamefont
  {Rogers}(2011)}]{Aybat:2011zv}%
  \BibitemOpen
  \bibfield  {author} {\bibinfo {author} {\bibfnamefont {S.~M.}\ \bibnamefont
  {Aybat}}\ and\ \bibinfo {author} {\bibfnamefont {T.~C.}\ \bibnamefont
  {Rogers}},\ }\href {\doibase 10.1103/PhysRevD.83.114042} {\bibfield
  {journal} {\bibinfo  {journal} {Phys.Rev.}\ }\textbf {\bibinfo {volume}
  {D83}},\ \bibinfo {pages} {114042} (\bibinfo {year} {2011})},\ \Eprint
  {http://arxiv.org/abs/1101.5057} {arXiv:1101.5057 [hep-ph]} \BibitemShut
  {NoStop}%
\bibitem [{\citenamefont {Aybat}\ \emph
  {et~al.}(2012{\natexlab{a}})\citenamefont {Aybat}, \citenamefont {Prokudin},\
  and\ \citenamefont {Rogers}}]{Aybat:2011ta}%
  \BibitemOpen
  \bibfield  {author} {\bibinfo {author} {\bibfnamefont {S.~M.}\ \bibnamefont
  {Aybat}}, \bibinfo {author} {\bibfnamefont {A.}~\bibnamefont {Prokudin}}, \
  and\ \bibinfo {author} {\bibfnamefont {T.~C.}\ \bibnamefont {Rogers}},\
  }\href {\doibase 10.1103/PhysRevLett.108.242003} {\bibfield  {journal}
  {\bibinfo  {journal} {Phys.Rev.Lett.}\ }\textbf {\bibinfo {volume} {108}},\
  \bibinfo {pages} {242003} (\bibinfo {year} {2012}{\natexlab{a}})},\ \Eprint
  {http://arxiv.org/abs/1112.4423} {arXiv:1112.4423 [hep-ph]} \BibitemShut
  {NoStop}%
\bibitem [{\citenamefont {Aybat}\ \emph
  {et~al.}(2012{\natexlab{b}})\citenamefont {Aybat}, \citenamefont {Collins},
  \citenamefont {Qiu},\ and\ \citenamefont {Rogers}}]{Aybat:2011ge}%
  \BibitemOpen
  \bibfield  {author} {\bibinfo {author} {\bibfnamefont {S.~M.}\ \bibnamefont
  {Aybat}}, \bibinfo {author} {\bibfnamefont {J.~C.}\ \bibnamefont {Collins}},
  \bibinfo {author} {\bibfnamefont {J.-W.}\ \bibnamefont {Qiu}}, \ and\
  \bibinfo {author} {\bibfnamefont {T.~C.}\ \bibnamefont {Rogers}},\ }\href
  {\doibase 10.1103/PhysRevD.85.034043} {\bibfield  {journal} {\bibinfo
  {journal} {Phys.Rev.}\ }\textbf {\bibinfo {volume} {D85}},\ \bibinfo {pages}
  {034043} (\bibinfo {year} {2012}{\natexlab{b}})},\ \Eprint
  {http://arxiv.org/abs/1110.6428} {arXiv:1110.6428 [hep-ph]} \BibitemShut
  {NoStop}%
\bibitem [{\citenamefont {Sun}\ and\ \citenamefont
  {Yuan}(2013{\natexlab{a}})}]{Sun:2013hua}%
  \BibitemOpen
  \bibfield  {author} {\bibinfo {author} {\bibfnamefont {P.}~\bibnamefont
  {Sun}}\ and\ \bibinfo {author} {\bibfnamefont {F.}~\bibnamefont {Yuan}},\
  }\href {\doibase 10.1103/PhysRevD.88.114012} {\bibfield  {journal} {\bibinfo
  {journal} {Phys.Rev.}\ }\textbf {\bibinfo {volume} {D88}},\ \bibinfo {pages}
  {114012} (\bibinfo {year} {2013}{\natexlab{a}})},\ \Eprint
  {http://arxiv.org/abs/1308.5003} {arXiv:1308.5003 [hep-ph]} \BibitemShut
  {NoStop}%
\bibitem [{\citenamefont {Sun}\ and\ \citenamefont
  {Yuan}(2013{\natexlab{b}})}]{Sun:2013dya}%
  \BibitemOpen
  \bibfield  {author} {\bibinfo {author} {\bibfnamefont {P.}~\bibnamefont
  {Sun}}\ and\ \bibinfo {author} {\bibfnamefont {F.}~\bibnamefont {Yuan}},\
  }\href {\doibase 10.1103/PhysRevD.88.034016} {\bibfield  {journal} {\bibinfo
  {journal} {Phys.Rev.}\ }\textbf {\bibinfo {volume} {D88}},\ \bibinfo {pages}
  {034016} (\bibinfo {year} {2013}{\natexlab{b}})},\ \Eprint
  {http://arxiv.org/abs/1304.5037} {arXiv:1304.5037 [hep-ph]} \BibitemShut
  {NoStop}%
\bibitem [{\citenamefont {Boer}(2013)}]{Boer:2013zca}%
  \BibitemOpen
  \bibfield  {author} {\bibinfo {author} {\bibfnamefont {D.}~\bibnamefont
  {Boer}},\ }\href {\doibase 10.1016/j.nuclphysb.2013.05.021} {\bibfield
  {journal} {\bibinfo  {journal} {Nucl.Phys.}\ }\textbf {\bibinfo {volume}
  {B874}},\ \bibinfo {pages} {217} (\bibinfo {year} {2013})},\ \Eprint
  {http://arxiv.org/abs/1304.5387} {arXiv:1304.5387 [hep-ph]} \BibitemShut
  {NoStop}%
\bibitem [{\citenamefont {Anselmino}\ \emph {et~al.}(2014)\citenamefont
  {Anselmino}, \citenamefont {Boglione}, \citenamefont {Gonzalez~H.},
  \citenamefont {Melis},\ and\ \citenamefont {Prokudin}}]{Anselmino:2013lza}%
  \BibitemOpen
  \bibfield  {author} {\bibinfo {author} {\bibfnamefont {M.}~\bibnamefont
  {Anselmino}}, \bibinfo {author} {\bibfnamefont {M.}~\bibnamefont {Boglione}},
  \bibinfo {author} {\bibfnamefont {J.}~\bibnamefont {Gonzalez~H.}}, \bibinfo
  {author} {\bibfnamefont {S.}~\bibnamefont {Melis}}, \ and\ \bibinfo {author}
  {\bibfnamefont {A.}~\bibnamefont {Prokudin}},\ }\href {\doibase
  10.1007/JHEP04(2014)005} {\bibfield  {journal} {\bibinfo  {journal} {JHEP}\
  }\textbf {\bibinfo {volume} {1404}},\ \bibinfo {pages} {005} (\bibinfo {year}
  {2014})},\ \Eprint {http://arxiv.org/abs/1312.6261} {arXiv:1312.6261
  [hep-ph]} \BibitemShut {NoStop}%
\bibitem [{\citenamefont {Anselmino}\ \emph
  {et~al.}(2013{\natexlab{a}})\citenamefont {Anselmino}, \citenamefont
  {Boglione}, \citenamefont {D'Alesio}, \citenamefont {Melis}, \citenamefont
  {Murgia} \emph {et~al.}}]{Anselmino:2013rya}%
  \BibitemOpen
  \bibfield  {author} {\bibinfo {author} {\bibfnamefont {M.}~\bibnamefont
  {Anselmino}}, \bibinfo {author} {\bibfnamefont {M.}~\bibnamefont {Boglione}},
  \bibinfo {author} {\bibfnamefont {U.}~\bibnamefont {D'Alesio}}, \bibinfo
  {author} {\bibfnamefont {S.}~\bibnamefont {Melis}}, \bibinfo {author}
  {\bibfnamefont {F.}~\bibnamefont {Murgia}},  \emph {et~al.},\ }\href
  {\doibase 10.1103/PhysRevD.88.054023} {\bibfield  {journal} {\bibinfo
  {journal} {Phys.Rev.}\ }\textbf {\bibinfo {volume} {D88}},\ \bibinfo {pages}
  {054023} (\bibinfo {year} {2013}{\natexlab{a}})},\ \Eprint
  {http://arxiv.org/abs/1304.7691} {arXiv:1304.7691 [hep-ph]} \BibitemShut
  {NoStop}%
\bibitem [{\citenamefont {Anselmino}\ \emph
  {et~al.}(2013{\natexlab{b}})\citenamefont {Anselmino}, \citenamefont
  {Boglione}, \citenamefont {D'Alesio}, \citenamefont {Melis}, \citenamefont
  {Murgia} \emph {et~al.}}]{Anselmino:2013vqa}%
  \BibitemOpen
  \bibfield  {author} {\bibinfo {author} {\bibfnamefont {M.}~\bibnamefont
  {Anselmino}}, \bibinfo {author} {\bibfnamefont {M.}~\bibnamefont {Boglione}},
  \bibinfo {author} {\bibfnamefont {U.}~\bibnamefont {D'Alesio}}, \bibinfo
  {author} {\bibfnamefont {S.}~\bibnamefont {Melis}}, \bibinfo {author}
  {\bibfnamefont {F.}~\bibnamefont {Murgia}},  \emph {et~al.},\ }\href
  {\doibase 10.1103/PhysRevD.87.094019} {\bibfield  {journal} {\bibinfo
  {journal} {Phys.Rev.}\ }\textbf {\bibinfo {volume} {D87}},\ \bibinfo {pages}
  {094019} (\bibinfo {year} {2013}{\natexlab{b}})},\ \Eprint
  {http://arxiv.org/abs/1303.3822} {arXiv:1303.3822 [hep-ph]} \BibitemShut
  {NoStop}%
\bibitem [{\citenamefont {Bacchetta}\ and\ \citenamefont
  {Prokudin}(2013)}]{Bacchetta:2013pqa}%
  \BibitemOpen
  \bibfield  {author} {\bibinfo {author} {\bibfnamefont {A.}~\bibnamefont
  {Bacchetta}}\ and\ \bibinfo {author} {\bibfnamefont {A.}~\bibnamefont
  {Prokudin}},\ }\href {\doibase 10.1016/j.nuclphysb.2013.07.013} {\bibfield
  {journal} {\bibinfo  {journal} {Nucl.Phys.}\ }\textbf {\bibinfo {volume}
  {B875}},\ \bibinfo {pages} {536} (\bibinfo {year} {2013})},\ \Eprint
  {http://arxiv.org/abs/1303.2129} {arXiv:1303.2129 [hep-ph]} \BibitemShut
  {NoStop}%
\bibitem [{\citenamefont {Aidala}\ \emph {et~al.}(2014)\citenamefont {Aidala},
  \citenamefont {Field}, \citenamefont {Gamberg},\ and\ \citenamefont
  {Rogers}}]{Aidala:2014hva}%
  \BibitemOpen
  \bibfield  {author} {\bibinfo {author} {\bibfnamefont {C.}~\bibnamefont
  {Aidala}}, \bibinfo {author} {\bibfnamefont {B.}~\bibnamefont {Field}},
  \bibinfo {author} {\bibfnamefont {L.}~\bibnamefont {Gamberg}}, \ and\
  \bibinfo {author} {\bibfnamefont {T.}~\bibnamefont {Rogers}},\ }\href
  {\doibase 10.1103/PhysRevD.89.094002} {\bibfield  {journal} {\bibinfo
  {journal} {Phys.Rev.}\ }\textbf {\bibinfo {volume} {D89}},\ \bibinfo {pages}
  {094002} (\bibinfo {year} {2014})},\ \Eprint {http://arxiv.org/abs/1401.2654}
  {arXiv:1401.2654 [hep-ph]} \BibitemShut {NoStop}%
\bibitem [{\citenamefont {Echevarria}\ \emph {et~al.}(2014)\citenamefont
  {Echevarria}, \citenamefont {Idilbi}, \citenamefont {Kang},\ and\
  \citenamefont {Vitev}}]{Echevarria:2014xaa}%
  \BibitemOpen
  \bibfield  {author} {\bibinfo {author} {\bibfnamefont {M.~G.}\ \bibnamefont
  {Echevarria}}, \bibinfo {author} {\bibfnamefont {A.}~\bibnamefont {Idilbi}},
  \bibinfo {author} {\bibfnamefont {Z.-B.}\ \bibnamefont {Kang}}, \ and\
  \bibinfo {author} {\bibfnamefont {I.}~\bibnamefont {Vitev}},\ }\href
  {\doibase 10.1103/PhysRevD.89.074013} {\bibfield  {journal} {\bibinfo
  {journal} {Phys.Rev.}\ }\textbf {\bibinfo {volume} {D89}},\ \bibinfo {pages}
  {074013} (\bibinfo {year} {2014})},\ \Eprint {http://arxiv.org/abs/1401.5078}
  {arXiv:1401.5078 [hep-ph]} \BibitemShut {NoStop}%
\bibitem [{\citenamefont {Echevarria}\ \emph
  {et~al.}(2013{\natexlab{b}})\citenamefont {Echevarria}, \citenamefont
  {Idilbi}, \citenamefont {Schafer},\ and\ \citenamefont
  {Scimemi}}]{Echevarria:2012pw}%
  \BibitemOpen
  \bibfield  {author} {\bibinfo {author} {\bibfnamefont {M.~G.}\ \bibnamefont
  {Echevarria}}, \bibinfo {author} {\bibfnamefont {A.}~\bibnamefont {Idilbi}},
  \bibinfo {author} {\bibfnamefont {A.}~\bibnamefont {Schafer}}, \ and\
  \bibinfo {author} {\bibfnamefont {I.}~\bibnamefont {Scimemi}},\ }\href
  {\doibase 10.1140/epjc/s10052-013-2636-y} {\bibfield  {journal} {\bibinfo
  {journal} {Eur.Phys.J.}\ }\textbf {\bibinfo {volume} {C73}},\ \bibinfo
  {pages} {2636} (\bibinfo {year} {2013}{\natexlab{b}})},\ \Eprint
  {http://arxiv.org/abs/1208.1281} {arXiv:1208.1281 [hep-ph]} \BibitemShut
  {NoStop}%
\bibitem [{\citenamefont {Echevarria}\ \emph {et~al.}(2012)\citenamefont
  {Echevarria}, \citenamefont {Idilbi},\ and\ \citenamefont
  {Scimemi}}]{GarciaEchevarria:2011rb}%
  \BibitemOpen
  \bibfield  {author} {\bibinfo {author} {\bibfnamefont {M.~G.}\ \bibnamefont
  {Echevarria}}, \bibinfo {author} {\bibfnamefont {A.}~\bibnamefont {Idilbi}},
  \ and\ \bibinfo {author} {\bibfnamefont {I.}~\bibnamefont {Scimemi}},\ }\href
  {\doibase 10.1007/JHEP07(2012)002} {\bibfield  {journal} {\bibinfo  {journal}
  {JHEP}\ }\textbf {\bibinfo {volume} {1207}},\ \bibinfo {pages} {002}
  (\bibinfo {year} {2012})},\ \Eprint {http://arxiv.org/abs/1111.4996}
  {arXiv:1111.4996 [hep-ph]} \BibitemShut {NoStop}%
\bibitem [{\citenamefont {Bauer}\ \emph {et~al.}(2001)\citenamefont {Bauer},
  \citenamefont {Fleming}, \citenamefont {Pirjol},\ and\ \citenamefont
  {Stewart}}]{Bauer:2000yr}%
  \BibitemOpen
  \bibfield  {author} {\bibinfo {author} {\bibfnamefont {C.~W.}\ \bibnamefont
  {Bauer}}, \bibinfo {author} {\bibfnamefont {S.}~\bibnamefont {Fleming}},
  \bibinfo {author} {\bibfnamefont {D.}~\bibnamefont {Pirjol}}, \ and\ \bibinfo
  {author} {\bibfnamefont {I.~W.}\ \bibnamefont {Stewart}},\ }\href {\doibase
  10.1103/PhysRevD.63.114020} {\bibfield  {journal} {\bibinfo  {journal}
  {Phys.Rev.}\ }\textbf {\bibinfo {volume} {D63}},\ \bibinfo {pages} {114020}
  (\bibinfo {year} {2001})},\ \Eprint {http://arxiv.org/abs/hep-ph/0011336}
  {arXiv:hep-ph/0011336 [hep-ph]} \BibitemShut {NoStop}%
\bibitem [{\citenamefont {Bauer}\ \emph {et~al.}(2002)\citenamefont {Bauer},
  \citenamefont {Pirjol},\ and\ \citenamefont {Stewart}}]{Bauer:2001yt}%
  \BibitemOpen
  \bibfield  {author} {\bibinfo {author} {\bibfnamefont {C.~W.}\ \bibnamefont
  {Bauer}}, \bibinfo {author} {\bibfnamefont {D.}~\bibnamefont {Pirjol}}, \
  and\ \bibinfo {author} {\bibfnamefont {I.~W.}\ \bibnamefont {Stewart}},\
  }\href {\doibase 10.1103/PhysRevD.65.054022} {\bibfield  {journal} {\bibinfo
  {journal} {Phys.Rev.}\ }\textbf {\bibinfo {volume} {D65}},\ \bibinfo {pages}
  {054022} (\bibinfo {year} {2002})},\ \Eprint
  {http://arxiv.org/abs/hep-ph/0109045} {arXiv:hep-ph/0109045 [hep-ph]}
  \BibitemShut {NoStop}%
\bibitem [{\citenamefont {Beneke}\ \emph {et~al.}(2002)\citenamefont {Beneke},
  \citenamefont {Chapovsky}, \citenamefont {Diehl},\ and\ \citenamefont
  {Feldmann}}]{Beneke:2002ph}%
  \BibitemOpen
  \bibfield  {author} {\bibinfo {author} {\bibfnamefont {M.}~\bibnamefont
  {Beneke}}, \bibinfo {author} {\bibfnamefont {A.}~\bibnamefont {Chapovsky}},
  \bibinfo {author} {\bibfnamefont {M.}~\bibnamefont {Diehl}}, \ and\ \bibinfo
  {author} {\bibfnamefont {T.}~\bibnamefont {Feldmann}},\ }\href {\doibase
  10.1016/S0550-3213(02)00687-9} {\bibfield  {journal} {\bibinfo  {journal}
  {Nucl.Phys.}\ }\textbf {\bibinfo {volume} {B643}},\ \bibinfo {pages} {431}
  (\bibinfo {year} {2002})},\ \Eprint {http://arxiv.org/abs/hep-ph/0206152}
  {arXiv:hep-ph/0206152 [hep-ph]} \BibitemShut {NoStop}%
\bibitem [{\citenamefont {Bauer}\ and\ \citenamefont
  {Stewart}(2001)}]{Bauer:2001ct}%
  \BibitemOpen
  \bibfield  {author} {\bibinfo {author} {\bibfnamefont {C.~W.}\ \bibnamefont
  {Bauer}}\ and\ \bibinfo {author} {\bibfnamefont {I.~W.}\ \bibnamefont
  {Stewart}},\ }\href {\doibase 10.1016/S0370-2693(01)00902-9} {\bibfield
  {journal} {\bibinfo  {journal} {Phys.Lett.}\ }\textbf {\bibinfo {volume}
  {B516}},\ \bibinfo {pages} {134} (\bibinfo {year} {2001})},\ \Eprint
  {http://arxiv.org/abs/hep-ph/0107001} {arXiv:hep-ph/0107001 [hep-ph]}
  \BibitemShut {NoStop}%
\bibitem [{\citenamefont {Idilbi}\ \emph {et~al.}(2004)\citenamefont {Idilbi},
  \citenamefont {Ji}, \citenamefont {Ma},\ and\ \citenamefont
  {Yuan}}]{Idilbi:2004vb}%
  \BibitemOpen
  \bibfield  {author} {\bibinfo {author} {\bibfnamefont {A.}~\bibnamefont
  {Idilbi}}, \bibinfo {author} {\bibfnamefont {X.-d.}\ \bibnamefont {Ji}},
  \bibinfo {author} {\bibfnamefont {J.-P.}\ \bibnamefont {Ma}}, \ and\ \bibinfo
  {author} {\bibfnamefont {F.}~\bibnamefont {Yuan}},\ }\href {\doibase
  10.1103/PhysRevD.70.074021} {\bibfield  {journal} {\bibinfo  {journal}
  {Phys.Rev.}\ }\textbf {\bibinfo {volume} {D70}},\ \bibinfo {pages} {074021}
  (\bibinfo {year} {2004})},\ \Eprint {http://arxiv.org/abs/hep-ph/0406302}
  {arXiv:hep-ph/0406302 [hep-ph]} \BibitemShut {NoStop}%
\bibitem [{\citenamefont {Barone}\ and\ \citenamefont
  {Ratcliffe}(2003)}]{Barone:2003fy}%
  \BibitemOpen
  \bibfield  {author} {\bibinfo {author} {\bibfnamefont {V.}~\bibnamefont
  {Barone}}\ and\ \bibinfo {author} {\bibfnamefont {P.}~\bibnamefont
  {Ratcliffe}},\ }\href@noop {} {\  (\bibinfo {year} {2003})}\BibitemShut
  {NoStop}%
\bibitem [{\citenamefont {Idilbi}\ \emph {et~al.}(2006)\citenamefont {Idilbi},
  \citenamefont {Ji},\ and\ \citenamefont {Yuan}}]{Idilbi:2006dg}%
  \BibitemOpen
  \bibfield  {author} {\bibinfo {author} {\bibfnamefont {A.}~\bibnamefont
  {Idilbi}}, \bibinfo {author} {\bibfnamefont {X.-d.}\ \bibnamefont {Ji}}, \
  and\ \bibinfo {author} {\bibfnamefont {F.}~\bibnamefont {Yuan}},\ }\href
  {\doibase 10.1016/j.nuclphysb.2006.07.002} {\bibfield  {journal} {\bibinfo
  {journal} {Nucl.Phys.}\ }\textbf {\bibinfo {volume} {B753}},\ \bibinfo
  {pages} {42} (\bibinfo {year} {2006})},\ \Eprint
  {http://arxiv.org/abs/hep-ph/0605068} {arXiv:hep-ph/0605068 [hep-ph]}
  \BibitemShut {NoStop}%
\bibitem [{\citenamefont {Idilbi}\ and\ \citenamefont
  {Scimemi}(2011)}]{Idilbi:2010im}%
  \BibitemOpen
  \bibfield  {author} {\bibinfo {author} {\bibfnamefont {A.}~\bibnamefont
  {Idilbi}}\ and\ \bibinfo {author} {\bibfnamefont {I.}~\bibnamefont
  {Scimemi}},\ }\href {\doibase 10.1016/j.physletb.2010.11.060} {\bibfield
  {journal} {\bibinfo  {journal} {Phys.Lett.}\ }\textbf {\bibinfo {volume}
  {B695}},\ \bibinfo {pages} {463} (\bibinfo {year} {2011})},\ \Eprint
  {http://arxiv.org/abs/1009.2776} {arXiv:1009.2776 [hep-ph]} \BibitemShut
  {NoStop}%
\bibitem [{\citenamefont {Garcia-Echevarria}\ \emph {et~al.}(2011)\citenamefont
  {Garcia-Echevarria}, \citenamefont {Idilbi},\ and\ \citenamefont
  {Scimemi}}]{GarciaEchevarria:2011md}%
  \BibitemOpen
  \bibfield  {author} {\bibinfo {author} {\bibfnamefont {M.}~\bibnamefont
  {Garcia-Echevarria}}, \bibinfo {author} {\bibfnamefont {A.}~\bibnamefont
  {Idilbi}}, \ and\ \bibinfo {author} {\bibfnamefont {I.}~\bibnamefont
  {Scimemi}},\ }\href {\doibase 10.1103/PhysRevD.84.011502} {\bibfield
  {journal} {\bibinfo  {journal} {Phys.Rev.}\ }\textbf {\bibinfo {volume}
  {D84}},\ \bibinfo {pages} {011502} (\bibinfo {year} {2011})},\ \Eprint
  {http://arxiv.org/abs/1104.0686} {arXiv:1104.0686 [hep-ph]} \BibitemShut
  {NoStop}%
\bibitem [{\citenamefont {Manohar}\ and\ \citenamefont
  {Stewart}(2007)}]{Manohar:2006nz}%
  \BibitemOpen
  \bibfield  {author} {\bibinfo {author} {\bibfnamefont {A.~V.}\ \bibnamefont
  {Manohar}}\ and\ \bibinfo {author} {\bibfnamefont {I.~W.}\ \bibnamefont
  {Stewart}},\ }\href {\doibase 10.1103/PhysRevD.76.074002} {\bibfield
  {journal} {\bibinfo  {journal} {Phys.Rev.}\ }\textbf {\bibinfo {volume}
  {D76}},\ \bibinfo {pages} {074002} (\bibinfo {year} {2007})},\ \Eprint
  {http://arxiv.org/abs/hep-ph/0605001} {arXiv:hep-ph/0605001 [hep-ph]}
  \BibitemShut {NoStop}%
\bibitem [{\citenamefont {Collins}\ and\ \citenamefont
  {Hautmann}(2000)}]{Collins:1999dz}%
  \BibitemOpen
  \bibfield  {author} {\bibinfo {author} {\bibfnamefont {J.~C.}\ \bibnamefont
  {Collins}}\ and\ \bibinfo {author} {\bibfnamefont {F.}~\bibnamefont
  {Hautmann}},\ }\href {\doibase 10.1016/S0370-2693(99)01384-2} {\bibfield
  {journal} {\bibinfo  {journal} {Phys.Lett.}\ }\textbf {\bibinfo {volume}
  {B472}},\ \bibinfo {pages} {129} (\bibinfo {year} {2000})},\ \Eprint
  {http://arxiv.org/abs/hep-ph/9908467} {arXiv:hep-ph/9908467 [hep-ph]}
  \BibitemShut {NoStop}%
\bibitem [{\citenamefont {Lee}\ and\ \citenamefont
  {Sterman}(2007)}]{Lee:2006nr}%
  \BibitemOpen
  \bibfield  {author} {\bibinfo {author} {\bibfnamefont {C.}~\bibnamefont
  {Lee}}\ and\ \bibinfo {author} {\bibfnamefont {G.~F.}\ \bibnamefont
  {Sterman}},\ }\href {\doibase 10.1103/PhysRevD.75.014022} {\bibfield
  {journal} {\bibinfo  {journal} {Phys.Rev.}\ }\textbf {\bibinfo {volume}
  {D75}},\ \bibinfo {pages} {014022} (\bibinfo {year} {2007})},\ \Eprint
  {http://arxiv.org/abs/hep-ph/0611061} {arXiv:hep-ph/0611061 [hep-ph]}
  \BibitemShut {NoStop}%
\bibitem [{\citenamefont {Idilbi}\ and\ \citenamefont
  {Mehen}(2007{\natexlab{a}})}]{Idilbi:2007ff}%
  \BibitemOpen
  \bibfield  {author} {\bibinfo {author} {\bibfnamefont {A.}~\bibnamefont
  {Idilbi}}\ and\ \bibinfo {author} {\bibfnamefont {T.}~\bibnamefont {Mehen}},\
  }\href {\doibase 10.1103/PhysRevD.75.114017} {\bibfield  {journal} {\bibinfo
  {journal} {Phys.Rev.}\ }\textbf {\bibinfo {volume} {D75}},\ \bibinfo {pages}
  {114017} (\bibinfo {year} {2007}{\natexlab{a}})},\ \Eprint
  {http://arxiv.org/abs/hep-ph/0702022} {arXiv:hep-ph/0702022 [HEP-PH]}
  \BibitemShut {NoStop}%
\bibitem [{\citenamefont {Idilbi}\ and\ \citenamefont
  {Mehen}(2007{\natexlab{b}})}]{Idilbi:2007yi}%
  \BibitemOpen
  \bibfield  {author} {\bibinfo {author} {\bibfnamefont {A.}~\bibnamefont
  {Idilbi}}\ and\ \bibinfo {author} {\bibfnamefont {T.}~\bibnamefont {Mehen}},\
  }\href {\doibase 10.1103/PhysRevD.76.094015} {\bibfield  {journal} {\bibinfo
  {journal} {Phys.Rev.}\ }\textbf {\bibinfo {volume} {D76}},\ \bibinfo {pages}
  {094015} (\bibinfo {year} {2007}{\natexlab{b}})},\ \Eprint
  {http://arxiv.org/abs/0707.1101} {arXiv:0707.1101 [hep-ph]} \BibitemShut
  {NoStop}%
\bibitem [{\citenamefont {Boer}\ \emph {et~al.}(2011)\citenamefont {Boer},
  \citenamefont {Gamberg}, \citenamefont {Musch},\ and\ \citenamefont
  {Prokudin}}]{Boer:2011xd}%
  \BibitemOpen
  \bibfield  {author} {\bibinfo {author} {\bibfnamefont {D.}~\bibnamefont
  {Boer}}, \bibinfo {author} {\bibfnamefont {L.}~\bibnamefont {Gamberg}},
  \bibinfo {author} {\bibfnamefont {B.}~\bibnamefont {Musch}}, \ and\ \bibinfo
  {author} {\bibfnamefont {A.}~\bibnamefont {Prokudin}},\ }\href {\doibase
  10.1007/JHEP10(2011)021} {\bibfield  {journal} {\bibinfo  {journal} {JHEP}\
  }\textbf {\bibinfo {volume} {1110}},\ \bibinfo {pages} {021} (\bibinfo {year}
  {2011})},\ \Eprint {http://arxiv.org/abs/1107.5294} {arXiv:1107.5294
  [hep-ph]} \BibitemShut {NoStop}%
\bibitem [{\citenamefont {Collins}\ and\ \citenamefont
  {Metz}(2004)}]{Collins:2004nx}%
  \BibitemOpen
  \bibfield  {author} {\bibinfo {author} {\bibfnamefont {J.~C.}\ \bibnamefont
  {Collins}}\ and\ \bibinfo {author} {\bibfnamefont {A.}~\bibnamefont {Metz}},\
  }\href {\doibase 10.1103/PhysRevLett.93.252001} {\bibfield  {journal}
  {\bibinfo  {journal} {Phys.Rev.Lett.}\ }\textbf {\bibinfo {volume} {93}},\
  \bibinfo {pages} {252001} (\bibinfo {year} {2004})},\ \Eprint
  {http://arxiv.org/abs/hep-ph/0408249} {arXiv:hep-ph/0408249 [hep-ph]}
  \BibitemShut {NoStop}%
\bibitem [{\citenamefont {Collins}\ \emph {et~al.}(1985)\citenamefont
  {Collins}, \citenamefont {Soper},\ and\ \citenamefont
  {Sterman}}]{Collins:1984kg}%
  \BibitemOpen
  \bibfield  {author} {\bibinfo {author} {\bibfnamefont {J.~C.}\ \bibnamefont
  {Collins}}, \bibinfo {author} {\bibfnamefont {D.~E.}\ \bibnamefont {Soper}},
  \ and\ \bibinfo {author} {\bibfnamefont {G.~F.}\ \bibnamefont {Sterman}},\
  }\href {\doibase 10.1016/0550-3213(85)90479-1} {\bibfield  {journal}
  {\bibinfo  {journal} {Nucl.Phys.}\ }\textbf {\bibinfo {volume} {B250}},\
  \bibinfo {pages} {199} (\bibinfo {year} {1985})}\BibitemShut {NoStop}%
\bibitem [{\citenamefont {Collins}(1993)}]{Collins:1992kk}%
  \BibitemOpen
  \bibfield  {author} {\bibinfo {author} {\bibfnamefont {J.~C.}\ \bibnamefont
  {Collins}},\ }\href {\doibase 10.1016/0550-3213(93)90262-N} {\bibfield
  {journal} {\bibinfo  {journal} {Nucl.Phys.}\ }\textbf {\bibinfo {volume}
  {B396}},\ \bibinfo {pages} {161} (\bibinfo {year} {1993})},\ \Eprint
  {http://arxiv.org/abs/hep-ph/9208213} {arXiv:hep-ph/9208213 [hep-ph]}
  \BibitemShut {NoStop}%
\bibitem [{\citenamefont {Bacchetta}\ \emph {et~al.}(2004)\citenamefont
  {Bacchetta}, \citenamefont {D'Alesio}, \citenamefont {Diehl},\ and\
  \citenamefont {Miller}}]{Bacchetta:2004jz}%
  \BibitemOpen
  \bibfield  {author} {\bibinfo {author} {\bibfnamefont {A.}~\bibnamefont
  {Bacchetta}}, \bibinfo {author} {\bibfnamefont {U.}~\bibnamefont {D'Alesio}},
  \bibinfo {author} {\bibfnamefont {M.}~\bibnamefont {Diehl}}, \ and\ \bibinfo
  {author} {\bibfnamefont {C.~A.}\ \bibnamefont {Miller}},\ }\href {\doibase
  10.1103/PhysRevD.70.117504} {\bibfield  {journal} {\bibinfo  {journal}
  {Phys.Rev.}\ }\textbf {\bibinfo {volume} {D70}},\ \bibinfo {pages} {117504}
  (\bibinfo {year} {2004})},\ \Eprint {http://arxiv.org/abs/hep-ph/0410050}
  {arXiv:hep-ph/0410050 [hep-ph]} \BibitemShut {NoStop}%
\bibitem [{\citenamefont {Anselmino}\ \emph {et~al.}(2012)\citenamefont
  {Anselmino}, \citenamefont {Boglione},\ and\ \citenamefont
  {Melis}}]{Anselmino:2012aa}%
  \BibitemOpen
  \bibfield  {author} {\bibinfo {author} {\bibfnamefont {M.}~\bibnamefont
  {Anselmino}}, \bibinfo {author} {\bibfnamefont {M.}~\bibnamefont {Boglione}},
  \ and\ \bibinfo {author} {\bibfnamefont {S.}~\bibnamefont {Melis}},\ }\href
  {\doibase 10.1103/PhysRevD.86.014028} {\bibfield  {journal} {\bibinfo
  {journal} {Phys.Rev.}\ }\textbf {\bibinfo {volume} {D86}},\ \bibinfo {pages}
  {014028} (\bibinfo {year} {2012})},\ \Eprint {http://arxiv.org/abs/1204.1239}
  {arXiv:1204.1239 [hep-ph]} \BibitemShut {NoStop}%
\bibitem [{\citenamefont {Collins}\ and\ \citenamefont
  {Soper}(1982)}]{Collins:1981uw}%
  \BibitemOpen
  \bibfield  {author} {\bibinfo {author} {\bibfnamefont {J.~C.}\ \bibnamefont
  {Collins}}\ and\ \bibinfo {author} {\bibfnamefont {D.~E.}\ \bibnamefont
  {Soper}},\ }\href {\doibase 10.1016/0550-3213(82)90021-9} {\bibfield
  {journal} {\bibinfo  {journal} {Nucl.Phys.}\ }\textbf {\bibinfo {volume}
  {B194}},\ \bibinfo {pages} {445} (\bibinfo {year} {1982})}\BibitemShut
  {NoStop}%
\bibitem [{\citenamefont {Collins}\ and\ \citenamefont
  {Rogers}(2013)}]{Collins:2012uy}%
  \BibitemOpen
  \bibfield  {author} {\bibinfo {author} {\bibfnamefont {J.~C.}\ \bibnamefont
  {Collins}}\ and\ \bibinfo {author} {\bibfnamefont {T.~C.}\ \bibnamefont
  {Rogers}},\ }\href {\doibase 10.1103/PhysRevD.87.034018} {\bibfield
  {journal} {\bibinfo  {journal} {Phys.Rev.}\ }\textbf {\bibinfo {volume}
  {D87}},\ \bibinfo {pages} {034018} (\bibinfo {year} {2013})},\ \Eprint
  {http://arxiv.org/abs/1210.2100} {arXiv:1210.2100 [hep-ph]} \BibitemShut
  {NoStop}%
\bibitem [{\citenamefont {Manohar}(2003)}]{Manohar:2003vb}%
  \BibitemOpen
  \bibfield  {author} {\bibinfo {author} {\bibfnamefont {A.~V.}\ \bibnamefont
  {Manohar}},\ }\href {\doibase 10.1103/PhysRevD.68.114019} {\bibfield
  {journal} {\bibinfo  {journal} {Phys.Rev.}\ }\textbf {\bibinfo {volume}
  {D68}},\ \bibinfo {pages} {114019} (\bibinfo {year} {2003})},\ \Eprint
  {http://arxiv.org/abs/hep-ph/0309176} {arXiv:hep-ph/0309176 [hep-ph]}
  \BibitemShut {NoStop}%
\bibitem [{\citenamefont {Ji}\ \emph {et~al.}(2005)\citenamefont {Ji},
  \citenamefont {Ma},\ and\ \citenamefont {Yuan}}]{Ji:2004wu}%
  \BibitemOpen
  \bibfield  {author} {\bibinfo {author} {\bibfnamefont {X.-d.}\ \bibnamefont
  {Ji}}, \bibinfo {author} {\bibfnamefont {J.-p.}\ \bibnamefont {Ma}}, \ and\
  \bibinfo {author} {\bibfnamefont {F.}~\bibnamefont {Yuan}},\ }\href {\doibase
  10.1103/PhysRevD.71.034005} {\bibfield  {journal} {\bibinfo  {journal}
  {Phys.Rev.}\ }\textbf {\bibinfo {volume} {D71}},\ \bibinfo {pages} {034005}
  (\bibinfo {year} {2005})},\ \Eprint {http://arxiv.org/abs/hep-ph/0404183}
  {arXiv:hep-ph/0404183 [hep-ph]} \BibitemShut {NoStop}%
\bibitem [{\citenamefont {Chen}\ \emph {et~al.}(2007)\citenamefont {Chen},
  \citenamefont {Idilbi},\ and\ \citenamefont {Ji}}]{Chen:2006vd}%
  \BibitemOpen
  \bibfield  {author} {\bibinfo {author} {\bibfnamefont {P.-y.}\ \bibnamefont
  {Chen}}, \bibinfo {author} {\bibfnamefont {A.}~\bibnamefont {Idilbi}}, \ and\
  \bibinfo {author} {\bibfnamefont {X.-d.}\ \bibnamefont {Ji}},\ }\href
  {\doibase 10.1016/j.nuclphysb.2006.11.020} {\bibfield  {journal} {\bibinfo
  {journal} {Nucl.Phys.}\ }\textbf {\bibinfo {volume} {B763}},\ \bibinfo
  {pages} {183} (\bibinfo {year} {2007})},\ \Eprint
  {http://arxiv.org/abs/hep-ph/0607003} {arXiv:hep-ph/0607003 [hep-ph]}
  \BibitemShut {NoStop}%
\bibitem [{\citenamefont {Kang}\ \emph {et~al.}(2011)\citenamefont {Kang},
  \citenamefont {Xiao},\ and\ \citenamefont {Yuan}}]{Kang:2011mr}%
  \BibitemOpen
  \bibfield  {author} {\bibinfo {author} {\bibfnamefont {Z.-B.}\ \bibnamefont
  {Kang}}, \bibinfo {author} {\bibfnamefont {B.-W.}\ \bibnamefont {Xiao}}, \
  and\ \bibinfo {author} {\bibfnamefont {F.}~\bibnamefont {Yuan}},\ }\href
  {\doibase 10.1103/PhysRevLett.107.152002} {\bibfield  {journal} {\bibinfo
  {journal} {Phys.Rev.Lett.}\ }\textbf {\bibinfo {volume} {107}},\ \bibinfo
  {pages} {152002} (\bibinfo {year} {2011})},\ \Eprint
  {http://arxiv.org/abs/1106.0266} {arXiv:1106.0266 [hep-ph]} \BibitemShut
  {NoStop}%
\bibitem [{\citenamefont {Ma}\ \emph {et~al.}(2013)\citenamefont {Ma},
  \citenamefont {Wang},\ and\ \citenamefont {Zhao}}]{Ma:2012hh}%
  \BibitemOpen
  \bibfield  {author} {\bibinfo {author} {\bibfnamefont {J.}~\bibnamefont
  {Ma}}, \bibinfo {author} {\bibfnamefont {J.}~\bibnamefont {Wang}}, \ and\
  \bibinfo {author} {\bibfnamefont {S.}~\bibnamefont {Zhao}},\ }\href {\doibase
  10.1103/PhysRevD.88.014027} {\bibfield  {journal} {\bibinfo  {journal}
  {Phys.Rev.}\ }\textbf {\bibinfo {volume} {D88}},\ \bibinfo {pages} {014027}
  (\bibinfo {year} {2013})},\ \Eprint {http://arxiv.org/abs/1211.7144}
  {arXiv:1211.7144 [hep-ph]} \BibitemShut {NoStop}%
\end{thebibliography}%

\end{document}